%% file: main.tex
\documentclass[11pt,a4paper]{article}

\makeatletter
\let\articlemakecaption\@makecaption
\makeatother
\usepackage{paper}
\makeatletter
\let\@makecaption\articlemakecaption
\makeatother
\usepackage[style=base]{subcaption}
\usepackage{mathrsfs}
\usepackage{dsfont}
\usepackage{placeins}
\usepackage{tikz}
\newsavebox{\leftcoefficienttable}
\newsavebox{\rightcoefficienttable}
\newlength{\coefficientpanelheight}
\newcommand{\setcoefficientpanelheight}{%
  \setlength{\coefficientpanelheight}{\dimexpr
    \ht\leftcoefficienttable+\dp\leftcoefficienttable\relax}%
  \ifdim\dimexpr\ht\rightcoefficienttable+\dp\rightcoefficienttable\relax
      >\coefficientpanelheight
    \setlength{\coefficientpanelheight}{\dimexpr
      \ht\rightcoefficienttable+\dp\rightcoefficienttable\relax}%
  \fi
}
\newcommand{\coefficientpanelbody}[1]{%
  \par\noindent\vbox to \coefficientpanelheight{%
    \hbox to \linewidth{\hfil\usebox{#1}\hfil}%
    \vfil}%
  \par
}
\usepackage[nameinlink,noabbrev]{cleveref}
\newcommand{\sectionrefstyle}[1]{%
  \crefname{#1}{\S}{\S}%
  \Crefname{#1}{\S}{\S}%
  \AddToHook{begindocument/end}{%
  \crefformat{#1}{##2\S##1##3}%
  \Crefformat{#1}{##2\S##1##3}%
  \crefmultiformat{#1}{##2\S##1##3}{ and ##2\S##1##3}%
    {, ##2\S##1##3}{, and ##2\S##1##3}%
  \Crefmultiformat{#1}{##2\S##1##3}{ and ##2\S##1##3}%
    {, ##2\S##1##3}{, and ##2\S##1##3}%
  }%
}
\sectionrefstyle{section}
\sectionrefstyle{subsection}
\sectionrefstyle{subsubsection}
\crefname{figure}{Figure}{Figures}
\Crefname{figure}{Figure}{Figures}
\crefname{table}{Table}{Tables}
\Crefname{table}{Table}{Tables}
\crefname{subtable}{Table}{Tables}
\Crefname{subtable}{Table}{Tables}
\crefformat{subtable}{#2Table~#1#3}
\Crefformat{subtable}{#2Table~#1#3}
\crefname{subfigure}{Figure}{Figures}
\Crefname{subfigure}{Figure}{Figures}
\crefformat{subfigure}{#2Figure~#1#3}
\Crefformat{subfigure}{#2Figure~#1#3}
\crefname{appendix}{Appendix}{Appendices}
\Crefname{appendix}{Appendix}{Appendices}
\crefname{subappendix}{Appendix}{Appendices}
\Crefname{subappendix}{Appendix}{Appendices}
\crefname{subsubappendix}{Appendix}{Appendices}
\Crefname{subsubappendix}{Appendix}{Appendices}
\AddToHook{cmd/appendix/after}{%
  \crefalias{section}{appendix}%
  \crefalias{subsection}{subappendix}%
  \crefalias{subsubsection}{subsubappendix}%
}
\crefname{equation}{}{}
\Crefname{equation}{}{}
\usepackage[
  backend=biber,
  style=alphabetic,
  maxalphanames=10,
  maxnames=10
]{biblatex}
\usepackage[final,right]{showlabels}
\makeatletter
\AddToHook{env/equation/after}{%
  \global\let\SL@labelname\relax
}
\makeatother

\AddToHook{env/subtable/begin}{\renewcommand{\showlabelsetlabel}[1]{}}
\AddToHook{env/subfigure/begin}{\renewcommand{\showlabelsetlabel}[1]{}}
\input{macros}

\title{Mutual information to all orders from modular flow: a replica-free approach}
\author[a]{Hewei Frederic Jia}
\affiliation[a]{Institute for Advanced Study, Tsinghua University, Beijing, 100084, China}
\emailAdd{heweifred@gmail.com}

\abstract{We derive an all-orders long-distance expansion of vacuum mutual information between two disjoint balls in generic
$d$-dimensional CFTs. The derivation
uses a novel radial sewing expansion of the joint vacuum
density matrix, together with a modular response formula for relative entropy and the geometric modular flow of a single ball, without analytic continuation in replica number. The expansion
coefficients are modular integrals of modular-flowed
vacuum correlators. Our formula provides a systematic treatment of arbitrary multi-point sectors, and recovers known results on two-point sector derived from replica methods. The full two-point sector resums into a modular integral of the
thermal partition function on $S^{d-1}$, with an effective temperature depending on ball configuration and modular time.

We compute the expansion explicitly in CFTs. In two-dimensional Ising CFT, using analytic reductions and multidimensional numerical integration, we obtain the first ten radial orders in the long-distance expansion, which involve sphere correlators with up to twenty spin insertions. The truncated expansion agrees well with lattice data at finite separation, and approximately satisfies crossing symmetry. Analytic calculations also reproduce known expansion coefficients for 2d free compact boson at self-dual radius and its current sector, as well as for free
scalars in higher dimensions.}

\begin{document}

\maketitle
\flushbottom

\section{Introduction and summary of results}
\label{sec:introduction}
Mutual information is a measure of correlations between subsystems. In QFT, for two disjoint spatial regions $A$ and $B$, their mutual information is defined by \footnote{In continuum QFT, mutual information is rigorously defined using Araki relative entropy on local operator algebras~\cite{Witten:2018zxz}. Here we work with density matrices defined formally from path integrals.}
\begin{equation}
  I(A:B) = S(\rho_{AB}\Vert\rho_{A} \otimes \rho_{B}) = S_A + S_B - S_{AB}.
  \label{eq:MI-def}
\end{equation}
Here $\rho_I$ is the reduced density matrix for a subregion $I$, $S_I=-\operatorname{Tr}\rho_I\log\rho_I$ is the entanglement entropy of the subregion, and the relative entropy is defined by
\begin{equation}
S(\rho\Vert\sigma)
=\operatorname{Tr}\left[\rho(\log\rho-\log\sigma)\right].
\label{eq:Srel-def}
\end{equation}
Entanglement entropy and mutual information find a variety of applications in QFT and holography; we refer to reviews~\cite{Nishioka:2018khk,Casini:2022rlv} and references therein for more general backgrounds.

In this work we study the vacuum mutual information between two disjoint spatial balls in a generic \CFT{d}. While hard to compute exactly for disjoint regions at finite distance, the mutual information obeys several general properties: i) it is non-negative and monotonically increases as distance decreases; ii) at short distance, it provides a geometric regularization of entanglement entropy~\cite{Casini:2015woa}, and probes universal quantities such as the Euler-anomaly coefficient for even $d$ and the sphere free energy for odd $d$~\cite{CHM11}; iii) at long distance, it is expected to admit an expansion organized by local operator data, viz. the operator spectrum and OPE coefficients, analogously to the OPE expansion of ordinary CFT correlators. The aim of this work is to provide a systematic treatment of the long-distance expansion for generic \CFT{d}.

A standard approach to compute mutual information in CFTs is the replica method \cite{Calabrese:2004eu}, which requires analytic continuation in replica number $N$. While the continuation remains challenging at finite distance in generic theories~\cite{Calabrese:2010he}, considerable progress has been made in the long-distance expansion, where replica twist operators are expanded in local operators of the replicated theory before taking the continuation~\cite{Calabrese:2010he,Cardy:2013nua,Chen:2017hbk}. In particular, the contribution to mutual information obtained by analytically continuing the two-copy sector, the sector where there are operator insertions on two of $N$ replicas, can be expressed as modular integrals involving vacuum two-point functions of the original CFT. This was first derived in~\cite{Agon:2015ftl} for the leading scalar primary two-point contribution, and recently generalized to incorporate descendants and spinning operators in~\cite{ACGP25,ABPV26} using a twist operator reconstruction ansatz. While derived through replica constructions, these results eventually only depend on vacuum correlators and the geometric modular flow of single ball~\cite{CHM11} in original CFT, without requiring explicit knowledge of finite-replica correlators. This hints at a more direct relation between modular flow and long-distance expansion of mutual information, separate from the replica construction. It is therefore natural to seek a derivation that bypasses both replica number continuation and twist operator reconstruction, and extends this structure to higher-point contributions.

We provide such a derivation in this work. We start by developing a novel radial sewing expansion for the vacuum density matrix of disjoint balls $\rho_{AB}$. This is derived by working in the conformal frame where the two balls are concentric, and then inserting a complete set of states in radial quantization in the Euclidean path integral that prepares $\rho_{AB}$. The vacuum contribution reproduces the reference product state $\rho_A\otimes\rho_B$ in~\cref{eq:MI-def}, while the non-vacuum contributions organize the long-distance perturbation of $\rho_{AB}$. A general modular response formula for relative entropy~\cref{eq:Srel-def} where $\rho$ is treated as perturbation of reference state $\sigma$ has been developed in~\cite{SU18}; see also~\cite{Lashkari:2018tjh}. Combining these with the geometric modular flow of single ball leads us to a replica-free derivation of the long-distance expansion of mutual information to all orders.

\subsection{Summary of results.}
\paragraph{General formula.} Our main general result is \cref{eq:mi-master}, which gives a systematic all-orders long-distance expansion of vacuum mutual information between two disjoint balls in generic \CFT{d}. The expansion is organized into $n$-point sectors. Each sector sums over dilatation eigenstates, and involves modular integrals of products of two modular-flowed vacuum $n$-point correlators, with universal kernels supplied by the modular response formula.

\paragraph{Resummation of two-point sector.} The full two-point sector contribution admits the compact representation \cref{eq:two-pt-sector-partition-function} as a modular integral of the thermal partition function of \CFT{d} on $S^{d-1}$, evaluated at an effective temperature determined by the ball configuration and modular time. Decomposing the partition function into conformal characters recovers the previously obtained long-module contributions~\cite{ACGP25,ABPV26}.

\paragraph{Relation with replica method.}Our~\cref{eq:mi-master} can also be understood from replica perspective: the $n$-point sector corresponds to $n$-copy sector, the sector where there are operator insertions on $n$ of the total $N$ replicas before continuation in $N$. The modular integrals can be identified as analytically-continued twist operator OPE coefficients, cf.~\cref{eq:C-equals-I}. In $d=2$ where twist operator OPE coefficients can be computed explicitly at finite $N$ from covering space method, the analytically continued OPE coefficients were generally obtained from numerical extrapolation~\cite{RTC18}. Our modular integrals therefore provide exact representations of these coefficients, and we find good agreement with numerical extrapolation (cf.~\cref{sec:replica-relation}).

\paragraph{Examples.} We study~\cref{eq:mi-master} in the following theories. The examples provide analytic checks against known results in free theories, as well as encouraging agreement with lattice data at finite separation in the Ising \CFT{2}.

\noindent\emph{Ising CFT$_2$} (\cref{sec:ising}): We compute the first ten radial orders, which involve correlators with
up to twenty spin insertions. The associated modular integrals
are evaluated using analytic reductions and quasi-Monte Carlo
integration. The truncated expansion agrees well with
lattice data at finite separation, and approximately satisfies crossing
symmetry of the non-universal part of mutual information in $d=2$.

\noindent\emph{Free compact boson CFT$_2$}
(\cref{sec:other-CFT2}): We analytically evaluate low-order modular-integral coefficients
in two cases: the full theory at the self-dual radius and the
radius-independent $U(1)$ current sector, reproducing known
results~\cite{CCT09,ACHP18}.
The current sector expansion is divergent~\cite{Abate:2025ywp},
whereas cancellations with vertex operator contributions produce
a convergent expansion for the full theory at self-dual radius.

\noindent\emph{Free scalar CFT$_d$ with $d\geq3$} (\cref{sec:scalar-CFTd}): We analytically evaluate low-order modular-integral coefficients
for $d=3,4$. The resulting mutual information expansion agrees
with~\cite{Buchanan:2024tlm}, which provides a systematic
replica-free method exploiting the Gaussian structure of the
theory. This gives an analytic check
of our general construction beyond $d=2$.

\paragraph{Organization.}
In~\cref{sec:relative-entropy-perturbation}, we review the modular
response formula for relative entropy.
In~\cref{sec:mi-expansion}, we derive the all-orders expansion
of mutual information and study its two- and three-point sectors.
We apply the framework to two-dimensional CFTs
in~\cref{sec:cft2-examples} and higher-dimensional free scalars
in~\cref{sec:scalar-CFTd}.
We discuss implications and future directions
in~\cref{sec:discussion}.
\Cref{app:ising-lattice} describes the lattice computation of mutual information for 2d Ising.

\section{Perturbation of relative entropy}
\label{sec:relative-entropy-perturbation}
We review a modular response formula for the perturbation of relative entropy~\cref{eq:Srel-def} when $\rho$ is treated as perturbation of reference state $\sigma$~\cite{SU18,Lashkari:2018tjh}. This prepares later application to mutual information~\cref{eq:MI-def} where $\rho_{AB}$ is treated as perturbation of $\rho_{A} \otimes \rho_{B}$ at long distance.

\subsection{Modular response formula}
\paragraph{Response coefficient from resolvent.} Let $\rho$ be a perturbation of reference state $\sigma$
\begin{equation}
\rho=\sigma+\delta\rho,
\qquad
\delta\rho^\dagger=\delta\rho,
\qquad
\operatorname{Tr}\delta\rho=0.
\end{equation}
The difference in operator logarithm admits integral representation using resolvent as
\begin{equation}
\log\rho-\log\sigma
=
\int_0^\infty d\lambda\,
\bigl[R_\sigma(\lambda)-R_\rho(\lambda)\bigr],
\end{equation}
where the resolvent is defined by
\begin{equation}
R_\rho(\lambda)=(\rho+\lambda\id)^{-1}.
\end{equation}
We expand the resolvent for $\rho$ in terms of the resolvent for reference state $\sigma$ as
\begin{equation}\label{eq:Rrho-expansion}
R_\rho(\lambda)
=
\sum_{k=0}^{\infty}
(-1)^k
\bigl[R_\sigma(\lambda)\delta\rho\bigr]^k
R_\sigma(\lambda).
\end{equation}
Substituting the resolvent expansion~\cref{eq:Rrho-expansion} into~\cref{eq:Srel-def} gives the formal expansion
\begin{equation}
S(\rho\Vert\sigma)
=\sum_{n=2}^{\infty}S^{[n]}(\delta\rho;\sigma).
\end{equation}
Here the response coefficient $S^{[n]}$ is of degree $n$ in $\delta\rho$:
\begin{equation}
\begin{aligned}
S^{[n]}(\delta\rho;\sigma)
&=
(-1)^n\int_0^\infty d\lambda\,
\operatorname{Tr}\!\left[
\bigl(R_\sigma(\lambda)\delta\rho\bigr)^n
-\sigma R_\sigma(\lambda)\bigl(R_\sigma(\lambda)\delta\rho\bigr)^n
\right]
\\
&=
(-1)^n\int_0^\infty d\lambda\,\lambda\,
\operatorname{Tr}\!\left[
R_\sigma(\lambda)\bigl(R_\sigma(\lambda)\delta\rho\bigr)^n
\right]
\\
&=
-\frac{(-1)^n}{n}
\int_0^\infty d\lambda\,\lambda\,\frac{d}{d\lambda}
\operatorname{Tr}\!\left[
\bigl(R_\sigma(\lambda)\delta\rho\bigr)^n
\right]
\\
&=
\frac{(-1)^n}{n}
\int_0^\infty d\lambda\,
\operatorname{Tr}\!\left[
\bigl(R_\sigma(\lambda)\delta\rho\bigr)^n
\right],
\end{aligned}
\label{eq:Sn-response}
\end{equation}
where we have used integration by parts in last equality. The $n=1$ response coefficient vanishes because
\begin{equation}
\begin{aligned}
S^{[1]}(\delta\rho;\sigma)
&=
\int_0^\infty d\lambda\,
\operatorname{Tr}\!\left[
\sigma R_\sigma(\lambda)\delta\rho R_\sigma(\lambda)
\right]
=
\operatorname{Tr}\!\left[
\delta\rho\int_0^\infty d\lambda\,
\sigma R_\sigma(\lambda)^2
\right]
=
\operatorname{Tr}\delta\rho
=0.
\end{aligned}
\end{equation}

\paragraph{Converting to modular integral.} We proceed to work with an orthonormal eigenbasis of the reference state
\begin{equation}
\sigma|r\rangle=p_r|r\rangle,
\qquad p_r>0.
\end{equation}
The resolvent factor in~\cref{eq:Sn-response} is then given by
\begin{equation}
\operatorname{Tr}\!\left[
\bigl(R_\sigma(\lambda)\delta\rho\bigr)^n
\right]
=
\sum_{r_0,\ldots,r_{n-1}}
\prod_{j=0}^{n-1}
\frac{\langle r_j|\delta\rho|r_{j+1}\rangle}
{p_{r_j}+\lambda},
\qquad r_n=r_0.
\end{equation}
We then have
\begin{equation}
S^{[n]}(\delta\rho;\sigma)
=\frac{(-1)^n}{n}
\sum_{r_0,\ldots,r_{n-1}}
\left[\prod_{j=0}^{n-1}\langle r_j|\delta\rho|r_{j+1}\rangle\right]
F_n(p_{r_0},\ldots,p_{r_{n-1}}), \qquad r_n=r_0,
\label{eq:Sn-matrix-element}
\end{equation}
with
\begin{equation}
F_n(p_0,\ldots,p_{n-1})
=\int_0^\infty d\lambda\,\prod_{j=0}^{n-1}(p_j+\lambda)^{-1}.
\end{equation}
We find it convenient to perform an overall rescaling to write
\begin{equation}
F_n(p_0,\ldots,p_{n-1})
=p_0^{1-n}f_n(p_1/p_0,\ldots,p_{n-1}/p_0), \qquad f_n(\boldsymbol q)
=\int_0^\infty\frac{du}{(1+u)\prod_{j=1}^{n-1}(q_j+u)},
\label{eq:fn-def}
\end{equation}
with $u = \lambda/p_0$ the scaled resolvent parameter. For each $1/(q+u)$ resolvent factor in the integrand of $f_n(\boldsymbol q)$, we use the integral representation
\begin{equation}
\frac1{q+u}
=
\frac1{4\pi}
\int_{\mathbb R+i\eta}dt\,
\frac{
q^{-1/2-it/(2\pi)}
u^{-1/2+it/(2\pi)}
}{
\cosh(t/2)
},
\qquad -\pi<\eta<\pi.
\label{eq:integral-representation}
\end{equation}
Here restriction to the strip $-\pi < \Im t < \pi$ is due to the poles of integrand at $t=\pm i \pi$. We then substitute \eqref{eq:integral-representation} into \eqref{eq:fn-def}, and use change of variable
\begin{equation}
  t_1=s_1,
\qquad
t_j=s_j-s_{j-1}-i\pi,
\quad j=2,\ldots,n-1
\label{eq:t-to-s}
\end{equation}
for each $1/(q_j+u)$ resolvent factor. This gives
\begin{equation}
\begin{aligned}
f_n(\boldsymbol q)
={}&
\frac{i^{n-2}}{(4\pi)^{n-1}}
\int_{\mathcal C_{n-1}}
\prod_{j=1}^{n-1}ds_j\,
\frac{
q_1^{-1/2-is_1/(2\pi)}
\displaystyle\prod_{j=2}^{n-1}
q_j^{-1+i(s_{j-1}-s_j)/(2\pi)}
}{
\cosh(s_1/2)
\displaystyle\prod_{j=2}^{n-1}
\sinh\!\left(\frac{s_j-s_{j-1}}2\right)
}
\\
&\times
\int_0^\infty
\frac{du\,u^{-1/2+is_{n-1}/(2\pi)}}{1+u}.
\end{aligned}
\label{eq:fn-u-integral}
\end{equation}
The integration contour $\mathcal C_{n-1}$ is determined as follows. The restriction to strip in~\cref{eq:integral-representation} and change of variable in~\cref{eq:t-to-s} gives the condition
\begin{equation}
-\pi<\Im s_1<\pi,
\qquad
0<\Im s_j-\Im s_{j-1}<2\pi,
\quad j\ge2.
\end{equation}
The convergence of resolvent parameter $u$ integral in~\cref{eq:fn-u-integral} additionally requires
\begin{equation}
  -\pi<\Im s_{n-1}<\pi.
\end{equation}
Together this gives the ordered contour
\begin{equation}
  \mathcal C_{n-1}
=
\prod_{j=1}^{n-1}
\left(
\mathbb R+i \theta_j
\right), \qquad -\pi<\theta_1<\cdots<\theta_{n-1}<\pi.
\label{eq:modular-integral-contour}
\end{equation}
Performing the $u$ integral in~\cref{eq:fn-u-integral} then yields
\begin{equation}
f_n(\boldsymbol q)
=\int_{\mathcal C_{n-1}}
\prod_{j=1}^{n-1}ds_j\,
\mathcal K_{n-1}(\boldsymbol s) \, q_1^{-1/2-is_1/(2\pi)}
\prod_{j=2}^{n-1}
q_j^{-1+i(s_{j-1}-s_j)/(2\pi)},
\label{eq:fn-modular-integral}
\end{equation}
with the modular kernel given by
\begin{equation}
  \mathcal K_{n-1}(\boldsymbol s)
=
\frac{i^{n-2}}{4^{n-1}\pi^{n-2}}
\frac{1}{
\cosh(s_1/2)\cosh(s_{n-1}/2)
\displaystyle\prod_{j=2}^{n-1}
\sinh\!\left(\frac{s_j-s_{j-1}}2\right)
}.
\end{equation}

We are now ready to obtain the modular response formula. From the spectral weights of $f_n$ in~\cref{eq:fn-modular-integral} and the rescaling in~\cref{eq:fn-def}, the spectral weights of $F_n$ in~\cref{eq:Sn-matrix-element} are
\begin{equation}
\begin{aligned}
&p_{r_0}^{1-n}
q_1^{-1/2-is_1/(2\pi)}
\prod_{j=2}^{n-1}
q_j^{-1+i(s_{j-1}-s_j)/(2\pi)}
=
\prod_{j=1}^{n-1}
p_{r_j}^{-1/2-is_j/(2\pi)}
p_{r_{j+1}}^{-1/2+is_j/(2\pi)}.
\end{aligned}
\end{equation}
This therefore combines with matrix elements in~\cref{eq:Sn-matrix-element} to yield
\begin{equation}
\begin{aligned}
&\sum_{r_0,\ldots,r_{n-1}}
\langle r_0|\delta\rho|r_1\rangle
\prod_{j=1}^{n-1}
\left[
p_{r_j}^{-1/2-is_j/(2\pi)}
\langle r_j|\delta\rho|r_{j+1}\rangle
p_{r_{j+1}}^{-1/2+is_j/(2\pi)}
\right]
\\
&=
\operatorname{Tr}\!\left[
\delta\rho
\prod_{j=1}^{n-1}
\left(
\sigma^{-1/2-is_j/(2\pi)}
\delta\rho\,
\sigma^{-1/2+is_j/(2\pi)}
\right)
\right].
\end{aligned}
\end{equation}
The response coefficient $S^{[n]}$ is then represented in terms of the following modular integral:
\begin{equation}
\boxed{
S^{[n]}(\delta\rho;\sigma)
=\frac{(-1)^n}{n}
\int_{\mathcal C_{n-1}}
\prod_{j=1}^{n-1}ds_j\,
\mathcal K_{n-1}(\boldsymbol s)
\,\operatorname{Tr}\left[
\delta\rho
\prod_{j=1}^{n-1}
\left(
\sigma^{-1/2-is_j/(2\pi)}\delta\rho
\sigma^{-1/2+is_j/(2\pi)}
\right)\right].
}
\label{eq:Srel-response}
\end{equation}
To make relation with modular flow more manifest, we may rewrite the integrand in~\cref{eq:Srel-response} as
\begin{equation}
  \operatorname{Tr}\left[
\delta\rho
\prod_{j=1}^{n-1}
\left(
\sigma^{-1/2-is_j/(2\pi)}\delta\rho
\sigma^{-1/2+is_j/(2\pi)}
\right)\right] = \operatorname{Tr}\!\left[
\sigma
\prod_{j=0}^{n-1}\delta\widetilde\rho(s_j)
\right], \quad s_0 = -i\pi,
\end{equation}
where
\begin{equation}
\delta\widetilde\rho
=
\sigma^{-1/2}\delta\rho\,\sigma^{-1/2},
\qquad
\delta\widetilde\rho(s)
=
\sigma^{-is/(2\pi)}
\delta\widetilde\rho\,
\sigma^{is/(2\pi)}.
\end{equation}
Thus the trace factor in~\cref{eq:Srel-response} is a correlation function of the sandwiched perturbation $\delta\widetilde\rho$ in the reference state $\sigma$, evaluated at complex modular times $s_j$. This representation is useful for later application to vacuum mutual information between disjoint balls in CFT: the reference state $\sigma=\rho_A\otimes\rho_B$ has geometric modular flow for each ball, while radial sewing expresses $\delta\widetilde\rho$ as a sum of products of local operator insertions.

\section{All-orders expansion of mutual information in \texorpdfstring{\CFT{d}}{CFTd}}
\label{sec:mi-expansion}

We now specialize to vacuum mutual information between two
disjoint balls in \CFT{d}. We first derive a radial sewing
expansion of their joint density matrix. Combining this expansion
with the geometric modular flow of single ball and the
modular response formula for relative entropy~\cref{eq:Srel-response},
we obtain the all-orders expansion~\cref{eq:mi-master} in terms
of modular integrals of vacuum CFT correlators. We then study
the two- and three-point sectors explicitly.

\subsection{Radial expansion for density matrix of disjoint balls}
\label{sec:radial-caps}
\paragraph{Radial expansion.}
We work in the conformal frame where two disjoint balls are concentric. We denote Euclidean space coordinate as $x = (\tau, \boldsymbol x) \in \mathbb{R}^d$. Let $A$ be the inner ball of radius $\xi$ and $B$ be the outer ball of unit radius:
\begin{equation}
A=\{\tau=0,\ |\boldsymbol x|<\xi\},
\qquad B=\{\tau=0,\ |\boldsymbol x|>1\},
\qquad 0<\xi<1.
\label{eq:regions}
\end{equation}
The radial parameter $\xi$ is related to the cross-ratio $\eta$ by\footnote{For a generic configuration of two balls with radii $R_A,R_B$ and center-to-center separation $L>R_A+R_B$, the cross-ratio is $
\eta
=
\frac{4R_AR_B}{L^2-(R_A-R_B)^2}
$.}
\begin{equation}
\eta=\frac{4\xi}{(1+\xi)^2}.
\label{eq:cross-ratio}
\end{equation}
The radial parameter $\xi$ coincides with the radial coordinate $\rho$ in context of conformal block~\cite{Hogervorst:2013sma}.

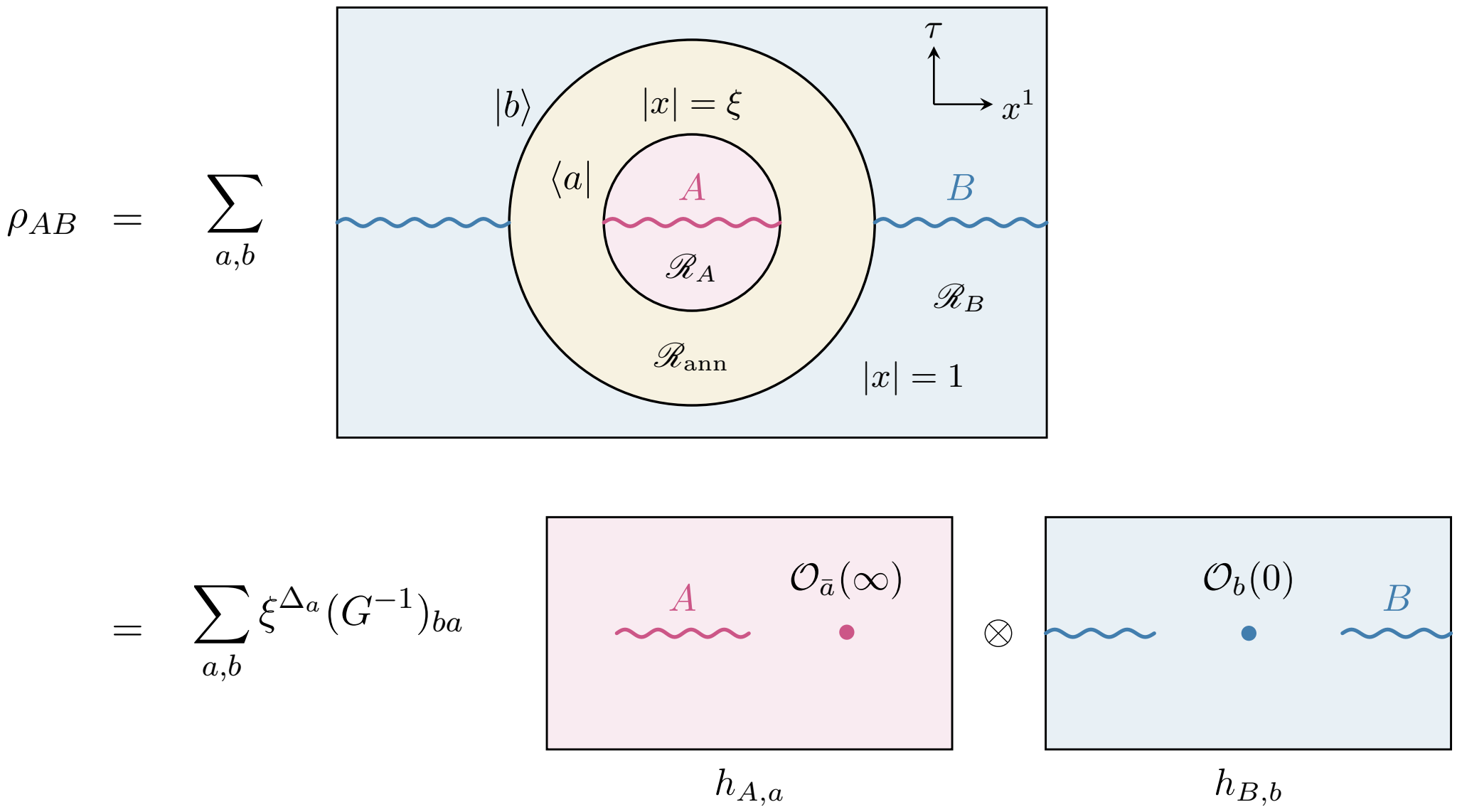
\begin{figure}[tbp]
\centering
\input{figures/radial-expansion}
\caption{The radial sewing expansion for density matrix $\rho_{AB}$ in~\cref{eq:radial-state}.}
\label{fig:radial-expansion}
\end{figure}

The vacuum density matrix $\rho_{AB}$ is prepared by Euclidean path integral on $\mathbb{R}^d$ with
$A$ and $B$ cuts~\cite{Calabrese:2004eu}. We decompose the path integral into the following three regions (cf. \cref{fig:radial-expansion}):
\begin{align}
\mathscr R_{A}&=\{x\in\mathbb R^d\mid |x|<\xi\}\setminus A,
\label{eq:inner-region}
\\
\mathscr R_{B}&=\{x\in\mathbb R^d\mid |x|>1\}\setminus B,
\label{eq:outer-region}
\\
\mathscr R_{\mathrm{ann}}&=\{x\in\mathbb R^d\mid \xi<|x|<1\}.
\label{eq:annulus}
\end{align}
The path integral in annulus region $\mathscr R_{\mathrm{ann}}$ gives the radial evolution operator
\begin{equation}
  \xi^{D}= \sum_{a,b}\xi^{\Delta_a}(G^{-1})_{ba}|b\rangle\langle a|
\end{equation}
where $D$ is dilatation operator, and the sum is over a complete basis of dilation eigenstates $|a\rangle$ with dimensions $\Delta_a$ on $\mathcal H_{S^{d-1}}$, with Gram matrix
\begin{equation}
  G_{ab}=\langle a|b\rangle.
  \label{eq:radial-gram}
\end{equation}
In other words, we insert $\langle a|$ at radius $|x|=\xi$ and $|b\rangle$ at radius $|x|=1$, and sum over $a,b$ in the path integral. Note that the radius here refers to the radius in radial quantization, not the radius of spatial balls $A$ and $B$.

Let $h_{A,a}$ be the overlap between the path integral on $\mathscr R_A$ and the state $\langle a|$ prepared at radius $\abs{x}=\xi$, and $h_{B,b}$ the overlap between the path integral on $\mathscr R_B$ and the state $|b\rangle$ prepared at radius $|x|=1$. By state-operator correspondence, the state $\langle a|$ at radius $\xi$ is prepared by the path integral on $\{x \in \mathbb{R}^d | \abs{x}>\xi\}$ with the following operator insertion at infinity:
\begin{equation}
\xi^{-\Delta_a}\,\mathcal O_{\bar a}(\infty).
\label{eq:radial-out}
\end{equation}
The insertion $\mathcal O_{\bar a}(\infty)$ prepares $\langle a|$ on the unit sphere, and the extra $\xi^{-\Delta_a}$ factor takes into account the radial evolution from $\abs{x}=1$ to $\abs{x}=\xi$. The conjugate operator $\mathcal O_{\bar a}(\infty)$ is defined from conjugation in radial quantization as follows. Radial conjugation reflects insertions through the unit sphere. In terms of conjugation in Euclidean time quantization, it takes the form
\begin{equation}
\mathcal O_a(x)^{\dagger_{\mathrm{rad}}}
=
U_{I\circ\Theta}\,
\mathcal O_a^\dagger(\Theta x)\,
U_{I\circ\Theta}^{-1},
\label{eq:radial-adjoint}
\end{equation}
where $U_f$ implements a conformal transformation $f$, $\Theta$ and $I$ denote Euclidean reflection and inversion
\begin{equation}
\Theta(\tau,\boldsymbol x)=(-\tau,\boldsymbol x),
\qquad I(x)=\frac{x}{x^2},
\label{eq:reflection-inversion}
\end{equation}
and $\dagger$ denotes conjugation in Euclidean time quantization. For example, for a scalar primary, the radial conjugation reduces to
\begin{equation}
\mathcal O(x)^{\dagger_{\mathrm{rad}}}
=(x^2)^{-\Delta}\,
\mathcal O^\dagger\!\left(\frac{x}{x^2}\right).
\end{equation}
The conjugate operator is then defined by
\begin{equation}
\mathcal O_{\bar a}(\infty)
:=\lim_{x\to0}\mathcal O_a(x)^{\dagger_{\mathrm{rad}}},
\label{eq:outgoing-insertion}
\end{equation}
and satisfies
\begin{equation}
  \left\langle\mathcal O_{\bar a}(\infty)\,\mathcal O_b(0)\right\rangle=\langle a|b\rangle = G_{ab}.
\label{eq:radial-dual}
\end{equation}
The path integral $h_{A,a}$ is therefore equivalent to path integral on $\mathbb{R}^d \setminus A$ with \cref{eq:radial-out} inserted at infinity. Similarly, the state $|b\rangle$ at unit radius is prepared by the path integral on $\{x \in \mathbb{R}^d | \abs{x}<1\}$ with operator
\begin{equation}
  \mathcal O_b(0)
\label{eq:radial-in}
\end{equation}
inserted at origin. The path integral $h_{B,b}$ is therefore equivalent to path integral on $\mathbb{R}^d \setminus B$ with \cref{eq:radial-in} inserted at origin.

In other words, $h_{A,a}$ and $h_{B,b}$ are the path integrals preparing $\rho_{A}$ and $\rho_{B}$ with one additional operator insertion, and we may write
\begin{equation}
\begin{aligned}
h_{A,a}
&=\xi^{-\Delta_a}\operatorname{Tr}_{A^c}
\left[|0\rangle\langle0|\,\mathcal O_{\bar a}(\infty)\right],\\
h_{B,b}
&=\operatorname{Tr}_{B^c}
\left[\mathcal O_b(0)|0\rangle\langle0|\right].
\end{aligned}
\label{eq:cap-prep}
\end{equation}
The radial expansion of the density matrix $\rho_{AB}$ is then given by
\begin{equation}
\rho_{AB}
=
\sum_{a,b}\xi^{\Delta_a}(G^{-1})_{ba}
\,h_{A,a}\otimes h_{B,b}.
\label{eq:radial-state}
\end{equation}
Since by definition
\begin{equation}
  h_{A,\id} = \rho_{A}, \quad h_{B,\id} = \rho_{B},
\end{equation}
the leading term in the long-distance expansion of $\rho_{AB}$ in \cref{eq:radial-state} is indeed the reference product state $\rho_A \otimes \rho_B$. We denote the perturbation to reference state as
\begin{equation}
  \delta\rho_{AB} := \rho_{AB} - \rho_A \otimes \rho_B = \sum_{a,b \neq \id}\xi^{\Delta_a}(G^{-1})_{ba}
\,h_{A,a}\otimes h_{B,b}.
\label{eq:delta-rhoAB}
\end{equation}
The perturbation density matrix~\cref{eq:delta-rhoAB} is not yet useful for direct application of the modular response formula~\cref{eq:Srel-response}. To make more progress, we need to use the geometric modular flow of single ball density matrices $\rho_{A},\rho_{B}$.

\paragraph{Geometric modular flow.}
Consider a spatial ball $\mathcal B_R$ of radius $R$, i.e., $\mathcal B_R = \{\{0,\boldsymbol x\} \in \mathbb{R}^d : |\boldsymbol x| < R\}$. Euclidean space with its entangling surface removed, $\mathbb{R}^d \setminus \partial \mathcal B_R$, is conformal to the modular cylinder $S^1\times H^{d-1}$ \cite{CHM11}:
\begin{equation}
\tau=\frac{R\sin\varphi}{\cosh u+\cos\varphi},\qquad
\boldsymbol x=\frac{R\sinh u}{\cosh u+\cos\varphi}\,\boldsymbol n,
\qquad \boldsymbol n^2=1,
\label{eq:ang-coords}
\end{equation}
\begin{equation}
ds_{\mathbb R^d}^2
=\frac{R^2}{(\cosh u+\cos\varphi)^2}
\left(d\varphi^2+du^2+\sinh^2u\,d\Omega_{d-2}^2\right).
\label{eq:ang-metric}
\end{equation}
The spatial ball corresponds to $\varphi=0$, and modular flow is rotation in modular angle $\varphi$:
\begin{equation}
g_\theta^{(R)}:(u,\varphi,\boldsymbol n)
\longmapsto(u,\varphi+\theta,\boldsymbol n),\qquad
\varphi\sim\varphi+2\pi.
\label{eq:angle-shift}
\end{equation}
Let $e_0=(1,\boldsymbol0)$ be unit vector in Euclidean time direction. The action of modular flow in $\mathbb{R}^d$ is given by
\begin{equation}
  g_\theta^{(R)}(x)
=\frac{x+\left[(\cos\theta-1)\tau
+\frac R2\sin\theta\left(1-\frac{x^2}{R^2}\right)\right]e_0}
{\cos^2(\theta/2)-\frac\tau R\sin\theta
+\frac{x^2}{R^2}\sin^2(\theta/2)}
\label{eq:ball-flow}
\end{equation}
with fixed point at entangling surface $\partial \mathcal{B}_R$.\footnote{For $d=2$, the spatial ball is interval $(-R,R)$, and modular cylinder is flat. The complex plane coordinate $z=x^1+i\tau$ is related to the complex modular cylinder coordinate $w=u+i\varphi$ by~\cite{CT16}
\begin{equation}
w=\log\frac{R+z}{R-z},
\qquad
z=R\tanh\frac w2.
\end{equation}
The action of modular flow in $z$ coordinate reads
\begin{equation}
g_\theta^{(R)}(z)
=
\frac{z\cos(\theta/2)+iR\sin(\theta/2)}
{\cos(\theta/2)+i(z/R)\sin(\theta/2)}
\label{eq:disk-flow}
\end{equation}
with fixed points at entangling points $z=\pm R$.
} In terms of conformal generators $P_\mu$ and $K_\mu$, the associated modular flow operator is given by
\begin{equation}
U_\theta^{(R)} := U\prn{g_\theta^{(R)}}
=\exp\!\left[\frac\theta2\left(RP_0+\frac{K_0}{R}\right)\right], \quad U_\theta^{(R)}|0\rangle=|0\rangle.
\label{eq:ball-gen}
\end{equation}
The vacuum density matrices of the ball and its spatial complement are related to the modular flow operator by
\begin{equation}
U_\theta^{(R)} = \rho_{\mathcal B_R}^{-\theta/(2\pi)}
\otimes\rho_{\mathcal B_R^c}^{\theta/(2\pi)}.
\label{eq:full-modular}
\end{equation}
We denote the modular-flowed local operator by
\begin{equation}
\mathcal O^{(\theta;R)}(x) := U_\theta^{(R)}\mathcal O(x)U_{-\theta}^{(R)}.
\label{eq:flowed-field}
\end{equation}
Superscript $R$ will be omitted for unit ball $R=1$. The modular flow operator transforms covariantly under dilatation:
\begin{equation}
U_\theta^{(R)}=R^D U_\theta R^{-D}.
\label{eq:mod-flow-dilatation-covariance}
\end{equation}

In particular, the action of modular flow by half-period is
\begin{equation}
g_\pi^{(R)}(x)=\frac{R^2\Theta x}{x^2},\qquad
g_\pi=I\circ\Theta.
\label{eq:half-period}
\end{equation}
Therefore the radial-conjugate operator \eqref{eq:outgoing-insertion} can be written in terms of modular flow operator as
\begin{equation}
\mathcal O_{\bar a}(\infty) = R^{2\Delta_a} U_\pi^{(R)}\mathcal O_a^\dagger(0)U_{-\pi}^{(R)}.
\label{eq:half-period-dual}
\end{equation}

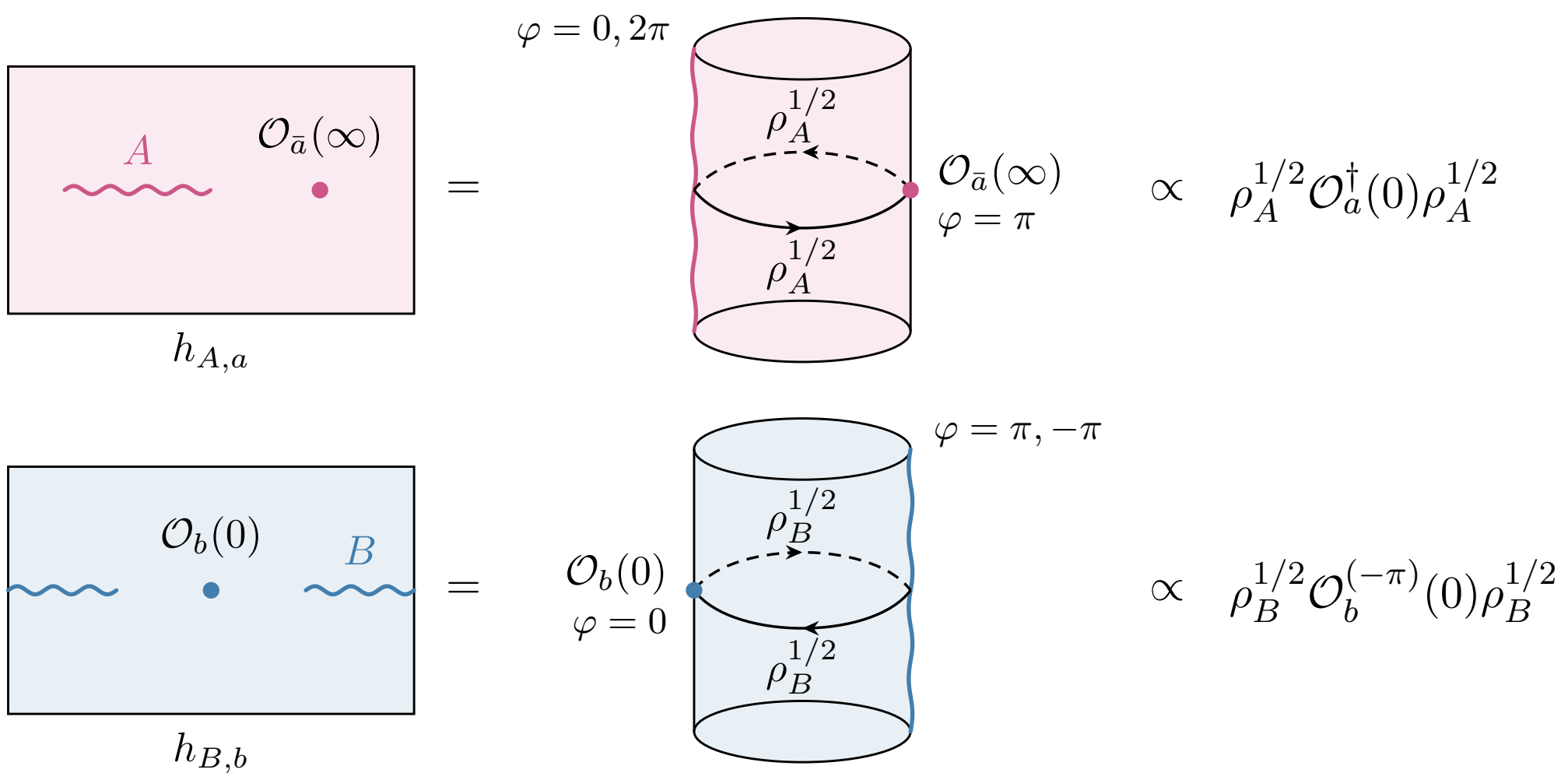
\begin{figure}[tbp]
\centering
\input{figures/modular-cap-preparation}
\caption{Radially expanded density matrices $h_{A,a}$ and $h_{B,b}$ in the sandwiched forms~\cref{eq:inner-cap,eq:outer-cap}. Modular flow by half-period in modular angle relates operator insertions at origin and infinity.}
\label{fig:modular-cap-preparation}
\end{figure}

\paragraph{Radial-expanded density matrices.} We may rewrite the operators inserted in radial-expanded density matrices \cref{eq:cap-prep} using modular flow operators as
\begin{equation}
\begin{aligned}
\mathcal O_{\bar a}(\infty)
&=
\xi^{2\Delta_a}
U_\pi^{(\xi)}\mathcal O_a^\dagger(0)U_{-\pi}^{(\xi)},\\
\mathcal O_b(0)
&=
U_\pi\mathcal O_b^{(-\pi)}(0)U_{-\pi},
\end{aligned}
\label{eq:cap-pullback}
\end{equation}
where in the first line we have used \cref{eq:half-period-dual}. Moreover, from \cref{eq:full-modular}, the modular operators are related to vacuum density matrices of $A,B$ and their complements by
\begin{equation}
  \begin{split}
    U^{(\xi)}_{\pi} &= \rho_A^{-1/2}\otimes\rho_{A^c}^{1/2}, \\
    U_{\pi} &= \rho_{B}^{1/2}\otimes\rho_{B^c}^{-1/2}.
  \end{split}
\label{eq:cap-modular}
\end{equation}
This allows us to write the radial-expanded density matrices in the following convenient forms:
\begin{align}
h_{A,a}
&=
\xi^{\Delta_a}
\operatorname{Tr}_{A^c}
\left[
|0\rangle\langle0|
U_\pi^{(\xi)}\mathcal O_a^\dagger(0)U_{-\pi}^{(\xi)}
\right]\notag\\
&=\xi^{\Delta_a}
\rho_A^{1/2}\mathcal O_a^\dagger(0)\rho_A^{1/2},
\label{eq:inner-cap}\\
h_{B,b}
&=
\operatorname{Tr}_{B^c}
\left[
U_\pi\mathcal O_b^{(-\pi)}(0)U_{-\pi}
|0\rangle\langle0|
\right]\notag\\
&=
\rho_B^{1/2}
\mathcal O_b^{(-\pi)}(0)
\rho_B^{1/2}.
\label{eq:outer-cap}
\end{align}
This is precisely of the sandwiched form discussed below~\cref{eq:Srel-response} that will facilitate application of the modular response formula. In CFT context, the half factor $\rho^{1/2}_{A,B}$ arises from using modular flow by half-period in modular angle to relate operator insertions at origin and infinity (cf.~\cref{fig:modular-cap-preparation}).

\subsection{The all-orders expansion}
\label{sec:mi-response}
We are now ready to apply the general modular response formula \eqref{eq:Srel-response} for perturbation of relative entropy to long-distance expansion of mutual information, with $\sigma=\rho_A\otimes\rho_B$ and $\delta\rho=\delta\rho_{AB}$ given by \cref{eq:delta-rhoAB}. Denoting the specialization of response coefficient $S^{[n]}$ in~\cref{eq:Srel-response} to mutual information as $I^{[n]}$, we have
\begin{align}
&I^{[n]}(\xi)=\frac{(-1)^n}{n}
\sum_{\substack{a_0,\ldots,a_{n-1}\ne\id\\
b_0,\ldots,b_{n-1}\ne\id}}
\xi^{\sum_j\Delta_{a_j}}
\prod_{j=0}^{n-1}(G^{-1})_{b_ja_j}
\int_{\mathcal C_{n-1}}\prod_{j=1}^{n-1}ds_j\,
\mathcal K_{n-1}(\boldsymbol s)
\label{eq:cap-response}
\\
&\quad\times\operatorname{Tr}_A\!\left[
h_{A,a_0}\prod_{j=1}^{n-1}
\left(
\rho_A^{-\frac12-\frac{is_j}{2\pi}}h_{A,a_j}
\rho_A^{-\frac12+\frac{is_j}{2\pi}}
\right)\right]
\operatorname{Tr}_B\!\left[
h_{B,b_0}\prod_{j=1}^{n-1}
\left(
\rho_B^{-\frac12-\frac{is_j}{2\pi}}h_{B,b_j}
\rho_B^{-\frac12+\frac{is_j}{2\pi}}
\right)\right].
\notag
\end{align}
The radial-expanded density matrices in the forms of \cref{eq:inner-cap,eq:outer-cap} allow us to rewrite the traces in terms of modular-flowed correlation functions. For the trace associated with the inner ball $A$, we have
\begin{equation}
\begin{aligned}
&\operatorname{Tr}_A\left[h_{A,a_0}\prod_{j=1}^{n-1}
\left(\rho_A^{-1/2-is_j/(2\pi)}h_{A,a_j}
\rho_A^{-1/2+is_j/(2\pi)}\right)\right]\\
&=\xi^{\sum_j\Delta_{a_j}}
\operatorname{Tr}_A\left[
\rho_A \prn{\rho^{-1/2}_A \mathcal O_{a_0}^\dagger(0) \rho^{1/2}_A }
\prod_{j=1}^{n-1}\left(
\rho_A^{-is_j/(2\pi)}\mathcal O_{a_j}^\dagger(0)
\rho_A^{is_j/(2\pi)}
\right)\right]\\
&=\xi^{\sum_j\Delta_{a_j}}
\left\langle\prod_{j=0}^{n-1}
(\mathcal O_{a_j}^\dagger)^{(is_j;\xi)}(0)\right\rangle, \qquad s_0=-i\pi\\
&=\left\langle\prod_{j=0}^{n-1}
(\mathcal O_{a_j}^\dagger)^{(is_j)}(0)\right\rangle.
\end{aligned}
\label{eq:inner-chain}
\end{equation}
In the last equality we have used dilatation covariance of modular flow \cref{eq:mod-flow-dilatation-covariance}. The resulting correlation functions therefore have operators modular-flowed using unit-ball modular Hamiltonian. For the trace associated with the outer ball $B$, we have
\begin{equation}
\begin{aligned}
&\operatorname{Tr}_B\left[h_{B,b_0}\prod_{j=1}^{n-1}
\left(\rho_B^{-1/2-is_j/(2\pi)}h_{B,b_j}
\rho_B^{-1/2+is_j/(2\pi)}\right)\right]\\
&=\operatorname{Tr}_B\left[
\rho_B \prn{\rho^{-1/2}_B \mathcal O_{b_0}^{(-\pi)}(0) \rho^{1/2}_B }
\prod_{j=1}^{n-1}\left(
\rho_B^{-is_j/(2\pi)}\mathcal O_{b_j}^{(-\pi)}(0)
\rho_B^{is_j/(2\pi)}
\right)\right]\\
&=\left\langle\mathcal O_{b_0}^{(-2\pi)}(0)
\prod_{j=1}^{n-1}\mathcal O_{b_j}^{(-\pi-is_j)}(0)\right\rangle\\
&=\left\langle\prod_{j=0}^{n-1}
\mathcal O_{b_j}^{(\pi-is_j)}(0)\right\rangle, \qquad s_0=-i\pi.
\end{aligned}
\label{eq:outer-chain}
\end{equation}
The last equality uses periodicity on modular cylinder.

We therefore arrive at the following all-orders long-distance expansion of mutual information in generic \CFT{d}:
\begin{equation}
  \boxed{
\begin{gathered}
I(A:B)\underset{\xi\to0}{\sim}
\sum_{n=2}^{\infty}I^{[n]}(\xi),\\[1ex]
\begin{aligned}
I^{[n]}(\xi)
={}&\frac{(-1)^n}{n}
\sum_{\substack{a_0,\ldots,a_{n-1}\ne\id\\
b_0,\ldots,b_{n-1}\ne\id}}
\xi^{\sum_{j=0}^{n-1}\Delta_{a_j}}
\prod_{j=0}^{n-1}(G^{-1})_{b_ja_j}\\
&\times\int_{\mathcal C_{n-1}}
\prod_{j=1}^{n-1}ds_j\,\mathcal K_{n-1}(\boldsymbol s)\left\langle\prod_{j=0}^{n-1}
(\mathcal O_{a_j}^\dagger)^{(is_j)}(0)\right\rangle
\left\langle\prod_{j=0}^{n-1}
\mathcal O_{b_j}^{(\pi-is_j)}(0)\right\rangle,
\qquad s_0=-i\pi.
\end{aligned}
\end{gathered}
  }
\label{eq:mi-master}
\end{equation}
This is our main general result. We recall the ingredients involved: $\xi$ is the radial parameter related to cross-ratio by~\cref{eq:cross-ratio}; the sum is over a complete basis of dilatation eigenstates, the conjugation $\dagger$ denotes conjugation in Euclidean time quantization, and $G_{ab}$ is the Gram matrix~\cref{eq:radial-gram}; the modular kernel is given by \cref{eq:fn-modular-integral}, and the integration contour by \cref{eq:modular-integral-contour}; the operators in vacuum correlators factors are modular-flowed via~\cref{eq:flowed-field} by the modular flow operator $U_{is}$ defined in~\cref{eq:ball-gen} at unit radius. The long-distance expansion is therefore organized as sum over $n$-point sectors, with each sector given by sum of modular integrals of modular-flowed $n$-point vacuum correlators. The dependence on the geometric configuration of the disjoint balls $A,B$ is through the radial parameter $\xi$ in the overall factor $\xi^{\sum_j\Delta_{a_j}}$.

\paragraph{Analytic continuation.} The correlator factors in~\cref{eq:mi-master} are analytically continued from the Euclidean correlators at purely imaginary modular times with the ordering in~\cref{eq:modular-integral-contour}. Consider the inner ball correlator factor as an example. The insertions with integrated modular times are modular-flowed from the origin by~\cref{eq:ball-flow}:
\begin{equation}
g_{is_j}(0)=i\tanh\frac{s_j}{2}\,e_0,
\qquad j=1,\ldots,n-1.
\end{equation}
Consider three-point function of Hermitian scalar primaries for concreteness. The modular-flowed correlator is given by
\begin{equation}\label{eq:three-pt-priamry-inner-ball}
\left\langle
\mathcal O_1^{(\pi)}(0)
\mathcal O_2^{(is_1)}(0)
\mathcal O_3^{(is_2)}(0)
\right\rangle
=
\frac{C_{123}}
{\left[\cosh(s_1/2)\right]^{\Delta_{12;3}}
 \left[\cosh(s_2/2)\right]^{\Delta_{13;2}}
 \left[-i\sinh((s_2-s_1)/2)\right]^{\Delta_{23;1}}},
\end{equation}
where $\Delta_{ij;k}=\Delta_i+\Delta_j-\Delta_k$, and $C_{123}$ is the three-point coefficient. The branches of fractional powers in the denominator are fixed by the ordering $-\pi< \Im s_1 < \Im s_2 < \pi$.

\paragraph{Scalar correlator factors.} The two correlators factors in~\cref{eq:mi-master} associated with inner and outer ball are distinct in general. They do coincide for a same pair of Hermitian scalar operators:
\begin{equation}
\begin{aligned}
&\left\langle
\mathcal O_0^{(\pi)}(0)
\prod_{j=1}^{n-1}\mathcal O_j^{(is_j)}(0)
\right\rangle
=
\left\langle
\mathcal O_0^{(-\pi)}(0)
\prod_{j=1}^{n-1}\mathcal O_j^{(-is_j)}(0)
\right\rangle
=
\left\langle
\mathcal O_0(0)
\prod_{j=1}^{n-1}\mathcal O_j^{(\pi-is_j)}(0)
\right\rangle .
\end{aligned}
\label{eq:factor-equality-scalar}
\end{equation}
The first equality follows from vacuum invariance under rotation in Euclidean plane $
R_{01}(\pi):
(\tau,x^1,\boldsymbol x_\perp)
\longmapsto
(-\tau,-x^1,\boldsymbol x_\perp)
$. This reverses the signs of $P_0$ and $K_0$, and therefore sends $U_\theta$ to $U_{-\theta}$. Scalar operators at the origin are invariant under this rotation; spinning operators instead acquire rotation matrices acting on their indices. The second equality follows from overall rotation in modular angle by $\pi$, which holds for generic operators.

\paragraph{Operator words grouping.} For practical applications of~\cref{eq:mi-master}, we group the non-vacuum states into mutually orthogonal Gram blocks of fixed scaling dimension, thereby having a manifestly basis-independent decomposition. An operator word $X = \{X_1,\dots, X_n \}$ denotes an unordered multiset of these blocks. For example, in Ising \CFT{2}, $X=\{\sigma,\sigma_1\}$ denotes one spin primary insertion and one insertion from the complete level-one subspace spanned by $\partial\sigma$ and $\bar\partial\sigma$, and contributes to $n=\abs{X}=2$ sector. The long distance expansion may then be organized as
\begin{equation}
I(\xi)\sim
\sum_{|X|\ge2}\xi^{\Delta_X}\mathcal I(X),
\label{eq:mi-X-sum}
\end{equation}
where
\begin{equation}
\begin{aligned}
\mathcal I(X)
:={}&
\frac{(-1)^n}{n}
\sum_{\boldsymbol\lambda\in\operatorname{Perm}(X)}
\sum_{\substack{a_j,b_j\in\lambda_j\\j=0,\ldots,n-1}}
\left[\prod_{j=0}^{n-1}(G^{-1})_{b_ja_j}\right]
\\
&\times
\int_{\mathcal C_{n-1}}
\prod_{j=1}^{n-1}ds_j\,
\mathcal K_{n-1}(\boldsymbol s)
\left\langle
\prod_{j=0}^{n-1}
(\mathcal O_{a_j}^{\dagger})^{(is_j)}(0)
\right\rangle
\left\langle
\prod_{j=0}^{n-1}
\mathcal O_{b_j}^{(\pi-is_j)}(0)
\right\rangle,
\end{aligned}
\label{eq:def-operator-word-coeff}
\end{equation}
where $n=|X|$ is the size of the operator word, $\Delta_X=\sum_{\lambda\in X}\Delta_\lambda$ is the total scaling dimension of the operator word, $\operatorname{Perm}(X)$ denotes distinct assignments of the blocks $\lambda_j$ to the $n$ insertion slots, and the sums over $a_j,b_j$ include all states within each assigned block.

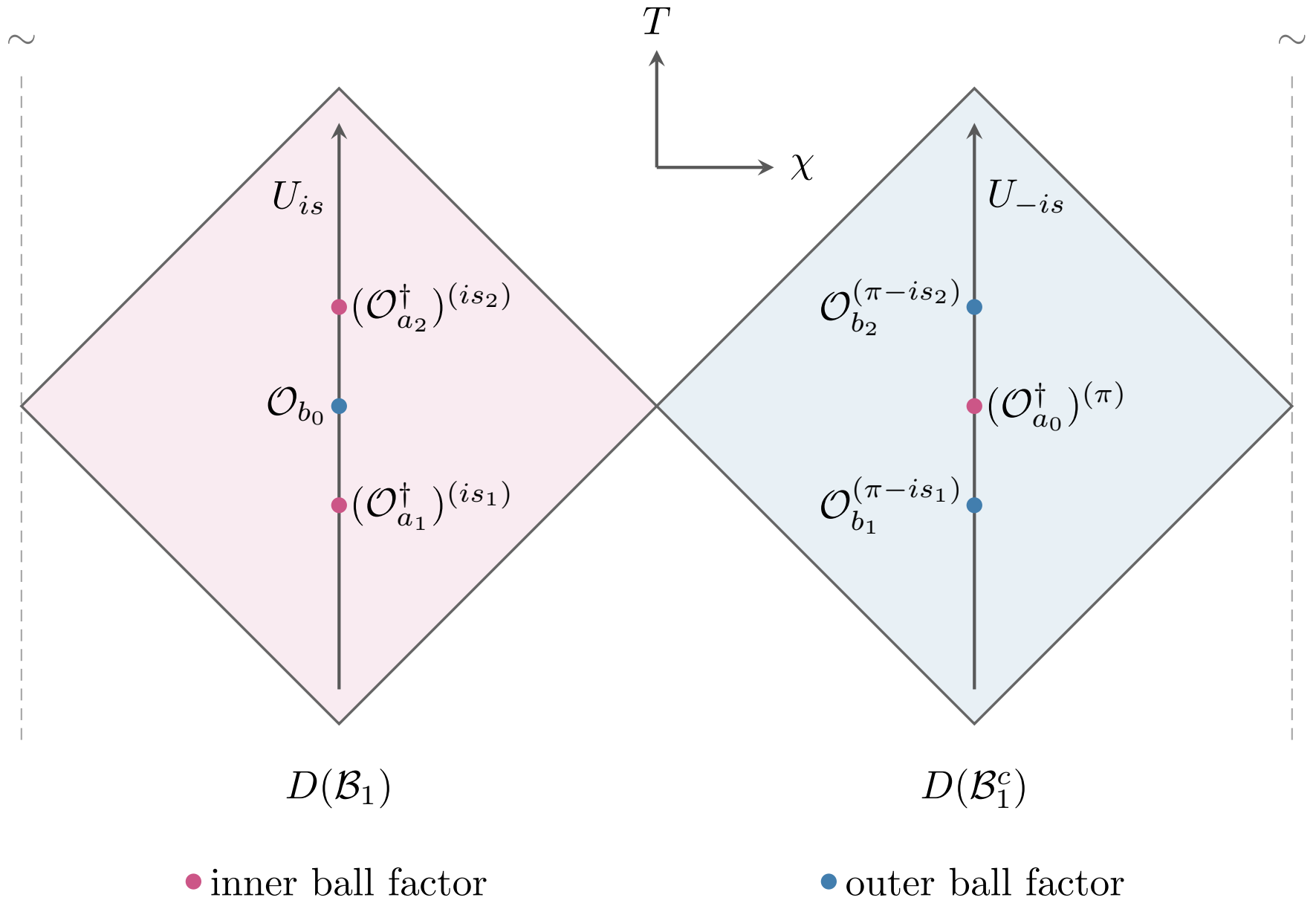
\begin{figure}[tbp]
\centering
\input{figures/lorentzian-modular-flow}
\caption{Lorentzian illustration of the $n=3$ integrand in \cref{eq:mi-master}. The diagram shows time $T$ and a spatial angle $\chi$ on the conformal cylinder $\mathbb R\times S^{d-1}$, with the remaining angular directions suppressed. Real modular trajectories are shown; the ordered imaginary shifts of the modular times prescribed in \cref{eq:modular-integral-contour} are understood. }
\label{fig:cubic-modular-flow}
\end{figure}

\subsection{Low-point sectors}
In the following we give explicit examples for $n=2,3$ sectors in~\cref{eq:mi-master}, where the relevant vacuum correlators are fixed by conformal symmetry. The two-point sector is studied in \cite{ACGP25,ABPV26}.
\subsubsection{Two-point sector}
The two-point sector contribution is
\begin{equation}
\begin{aligned}
I^{[2]}(\xi)
={}&\frac18
\sum_{\substack{a_0,a_1\ne\mathbf1\\b_0,b_1\ne\mathbf1}}
\xi^{\Delta_{a_0}+\Delta_{a_1}}
(G^{-1})_{b_0a_0}(G^{-1})_{b_1a_1}
\\
&\times\int_{\mathbb R}\frac{ds}{\cosh^2(s/2)}
\left\langle
(\mathcal O_{a_0}^{\dagger})^{(\pi)}(0)
(\mathcal O_{a_1}^{\dagger})^{(is)}(0)
\right\rangle
\left\langle
\mathcal O_{b_0}(0)
\mathcal O_{b_1}^{(\pi-is)}(0)
\right\rangle .
\end{aligned}
\label{eq:two-pt-sector}
\end{equation}

\paragraph{Scalar primary contribution.} For a unit-normalized Hermitian scalar primary $\mathcal O$ with dimension $\Delta$, its associated inner ball factor is
\begin{equation}
\left\langle
\mathcal O^{(\pi)}(0)\mathcal O^{(is)}(0)
\right\rangle
=
\frac{1}{\cosh^{2\Delta}(s/2)}.
\end{equation}
The outer ball factor is identical as explained in~\cref{eq:factor-equality-scalar}. The contribution to mutual information is
\begin{equation}
\begin{aligned}
\left.I^{[2]}(\xi)\right|_{\mathcal O}
&=
\frac{\xi^{2\Delta}}8
\int_{\mathbb R}
\frac{ds}{\cosh^{4\Delta+2}(s/2)} =
\frac{\sqrt\pi\,\Gamma(2\Delta+1)}
{4\Gamma(2\Delta+\tfrac32)}
\,\xi^{2\Delta}.
\end{aligned}
\label{eq:two-pt-sector-primary}
\end{equation}
This agrees with \cite{Calabrese:2010he,Agon:2015ftl}. In notation of~\cref{eq:mi-X-sum}, this is the contribution associated with operator word $X=\{\mathcal{O},\mathcal{O}\}$.
\paragraph{Full contribution.} Assuming we can exchange the order of modular integral and the sum over states in~\cref{eq:two-pt-sector}, the integrand may be resummed into a partition function. To start with, we may write the inner and outer ball factors as
\begin{equation}
\begin{aligned}
\left\langle
(\mathcal O_{a_0}^\dagger)^{(\pi)}(0)
(\mathcal O_{a_1}^\dagger)^{(is)}(0)
\right\rangle
&=\langle a_0|U_{is}|a_1^\dagger\rangle,
\\
\left\langle
\mathcal O_{b_0}(0)
\mathcal O_{b_1}^{(\pi-is)}(0)
\right\rangle
&=\langle b_1^\dagger|U_{is}|b_0\rangle,
\end{aligned}
\end{equation}
where the conjugate basis satisfies $
\langle a^\dagger|b^\dagger\rangle
=\overline{G_{ab}}=G_{ba}$. From completeness relation
\begin{equation}
\begin{aligned}
\sum_{a_0,b_0\ne\mathbf1}
\xi^{\Delta_{a_0}}(G^{-1})_{b_0a_0}
|b_0\rangle\langle a_0|
&=\xi^D-|0\rangle\langle0|,
\\
\sum_{a_1,b_1\ne\mathbf1}
\xi^{\Delta_{a_1}}(G^{-1})_{b_1a_1}
|a_1^\dagger\rangle\langle b_1^\dagger|
&=\xi^D-|0\rangle\langle0|,
\end{aligned}
\end{equation}
the integrand becomes
\begin{equation}
\operatorname{Tr}_{\mathcal H_{S^{d-1}}}
\!\left[
(\xi^D-|0\rangle\langle0|)U_{is}
(\xi^D-|0\rangle\langle0|)U_{is}
\right]
=
\operatorname{Tr}_{\mathcal H_{S^{d-1}}}
\!\left(\xi^D U_{is}\xi^D U_{is}\right)-1.
\end{equation}
Using conformal subalgebra
\begin{equation}
[D,P_0]=P_0,\qquad
[D,K_0]=-K_0,\qquad
[K_0,P_0]=2D
\end{equation}
and the definition of modular operator~\cref{eq:ball-gen}, it can be shown that the operator $\xi^D U_{is}$ is conjugate to a radial evolution operator
\begin{equation}
\xi^D U_{is}
=
V_{\xi,s}e^{-\beta(\xi,s)D/2}V_{\xi,s}^{-1},
\end{equation}
where
\begin{equation}
\begin{aligned}
\beta(\xi,s)
&=
4\operatorname{arccosh}\!\left[
\frac{1+\xi}{2\sqrt\xi}\cosh\frac s2
\right],
\\
\vartheta(\xi,s)
&=
\arctan\!\left[
\frac{2\sqrt\xi}{1-\xi}\tanh\frac s2
\right],
\\
V_{\xi,s}
&=
\xi^{D/2}
\exp\!\left[\frac{i\vartheta(\xi,s)}2(P_0-K_0)\right].
\end{aligned}
\end{equation}
From cyclicity of trace, the full two-point sector contribution can then be written as modular integral of the thermal partition function of \CFT{d}, $Z_{S^{d-1}}(\beta) = \operatorname{Tr}_{\mathcal H_{S^{d-1}}} e^{-\beta D}$, with an effective inverse temperature $\beta(\xi,s)$:
\begin{equation}
  \boxed{
\begin{aligned}
I^{[2]}(\xi)
=
\frac18\int_{\mathbb R}
\frac{ds}{\cosh^2(s/2)}
\left[
Z_{S^{d-1}}\!\left(\beta(\xi,s)\right)-1
\right].
\end{aligned}
  }
  \label{eq:two-pt-sector-partition-function}
\end{equation}
For a long (global) conformal module $\mathcal V_{\Delta, \mathcal{R}}$ generated by a primary of dimension $\Delta$ with $SO(d)$ representation $\mathcal R$, whose character is given by
\begin{equation}
\chi_{\Delta,\mathcal R}^{\mathrm{long}}(e^{-\beta})
=
\operatorname{Tr}_{\mathcal V_{\Delta,\mathcal R}}
e^{-\beta D}
=
(\dim\mathcal R)\,
\frac{e^{-\beta\Delta}}{(1-e^{-\beta})^d},
\end{equation}
its contribution to~\cref{eq:two-pt-sector-partition-function} can be evaluated in closed-form as
\begin{equation}
\begin{aligned}
\left.I^{[2]}(\xi)\right|_{\mathcal V_{\Delta,\mathcal R}}
={}&
(\dim\mathcal R)\,
\frac{\sqrt{\pi}\,\Gamma(2\Delta+1)}
{4\Gamma(2\Delta+\tfrac32)}
(1+\xi)^2\xi^{2\Delta}
\\
&\times
F_D^{(3)}\!\left(
2\Delta+1;\,
d-1,d+2,\tfrac12;\,
2\Delta+\tfrac32;\,
\xi,-\xi,\xi^2
\right).
\end{aligned}
\end{equation}
Here $F_D^{(3)}$ denotes Lauricella’s hypergeometric function of three variables. It can be verified that this agrees with the integral representation given in~\cite{ACGP25,ABPV26}. Its leading term in $\xi\to0$ limit indeed recovers~\cref{eq:two-pt-sector-primary}.

\subsubsection{Three-point sector}
The three-point sector contribution is
\begin{equation}
\begin{aligned}
I^{[3]}(\xi)
={}&-\frac13
\sum_{\substack{a_0,a_1,a_2\ne\mathbf1\\
b_0,b_1,b_2\ne\mathbf1}}
\xi^{\Delta_{a_0}+\Delta_{a_1}+\Delta_{a_2}}
\prod_{j=0}^{2}(G^{-1})_{b_ja_j}
\int_{\mathcal C_2}ds_1\,ds_2\,
\mathcal K_2(s_1,s_2)
\\
&\times
\left\langle
(\mathcal O_{a_0}^{\dagger})^{(\pi)}(0)
(\mathcal O_{a_1}^{\dagger})^{(is_1)}(0)
(\mathcal O_{a_2}^{\dagger})^{(is_2)}(0)
\right\rangle
\left\langle
\mathcal O_{b_0}(0)
\mathcal O_{b_1}^{(\pi-is_1)}(0)
\mathcal O_{b_2}^{(\pi-is_2)}(0)
\right\rangle .
\end{aligned}
\label{eq: three-pt-sector-primary}
\end{equation}

\paragraph{Scalar primary three-point contribution.} For Hermitian scalar primaries, the inner ball factor is given by~\cref{eq:three-pt-priamry-inner-ball}.
The outer ball factor is again identical as explained in~\cref{eq:factor-equality-scalar}. The identity we need for the resulting modular integral is
\begin{equation}
\begin{aligned}
&\int_{\mathcal C_2}ds_1\,ds_2\,
\frac{\mathcal K_2(s_1,s_2)}
{\cosh^{2\alpha}(s_1/2)
 \cosh^{2\beta}(s_2/2)
 \left[-\sinh^2((s_2-s_1)/2)\right]^\gamma}
\\
&\qquad=
\frac{\sqrt\pi}{4}\,
\frac{
\Gamma(\alpha+\beta+1)
\Gamma(\alpha+\gamma+1)
\Gamma(\beta+\gamma+1)
}{
\Gamma(\alpha+\beta+\gamma+\tfrac32)
\Gamma(\alpha+1)\Gamma(\beta+1)\Gamma(\gamma+1)
}.
\end{aligned}
\end{equation}
The identity holds in the domain $\operatorname{Re}(\alpha+\beta)>-1, \, \operatorname{Re}(\alpha+\gamma)>-1, \,\operatorname{Re}(\beta+\gamma)>-1$. We then obtain the three-point primary contribution as
\begin{equation}
  \boxed{
\begin{aligned}
\left.I^{[3]}(\xi)\right|_{\{\mathcal O_1,\mathcal O_2,\mathcal O_3\}}
={}&-\frac{\sqrt\pi}{2}\,
N_{123}C_{123}^{\,2}\,
\xi^{\Delta_{123}}
\\
&\times
\frac{
\displaystyle\prod_{i=1}^{3}\Gamma(2\Delta_i+1)
}{
\Gamma(\Delta_{123}+\tfrac32)
\Gamma(1+\Delta_{12;3})
\Gamma(1+\Delta_{13;2})
\Gamma(1+\Delta_{23;1})
},
\end{aligned}
  }
\label{eq: three-pt-sector-primary-result}
\end{equation}
where $\Delta_{123} = \Delta_{1} + \Delta_{2} + \Delta_{3}$, and the symmetry factor $N_{123}$ equals $1$ for three distinct operators, $1/2$ for two identical operators, and $1/6$ for three identical operators. In notation of~\cref{eq:mi-X-sum}, this is the contribution associated with operator word $X = \{\mathcal{O}_1,\mathcal{O}_2,\mathcal{O}_3 \}$. For three identical scalar primaries, this reproduces the result in~\cite{Takayanagi:2018zqx}.

\section{Examples in \CFT{2}}\label{sec:cft2-examples}
In this section we apply~\cref{eq:mi-master} to the Ising and
free compact boson \CFT{2}. Entanglement entropy of disjoint intervals in these theories has been
studied extensively, including exact integer-replica results
and long-distance expansions~\cite{CCT09,Calabrese:2010he},
lattice and tensor-network calculations~\cite{ATC10,Fagotti:2010yr},
and numerical replica continuation and conformal-block
expansions~\cite{DeNobili:2015dla,RTC18}.

For two disjoint intervals, it is
convenient to separate the mutual information into a universal term fixed
by the central charge and a theory-dependent function~\cite{CCT09}:
\begin{equation}
I(\eta)=-\frac c3\log(1-\eta)+F(\eta).
\label{eq:mi-crossing-function}
\end{equation}
In $d=2$, exchanging the two intervals with their complement
sends $\eta\to1-\eta$. For the full theories considered here, vacuum purity
and the universal single-interval entropy imply
\begin{equation}
F(\eta)=F(1-\eta).
\label{eq:mi-crossing-symmetry}
\end{equation}
For $d>2$, the complement of two disjoint balls is connected,
so there is no analogous crossing relation for mutual information at higher dimensions.
The crossing symmetry~\cref{eq:mi-crossing-symmetry} provides a non-perturbative diagnostic for our long-distance expansion.

\subsection{Ising}\label{sec:ising}
The Ising \CFT{2} has central charge $c=\bar{c}=1/2$; its Hilbert space has decomposition
\begin{equation}
\mathcal H_{\mathrm{Ising}}
=
\bigoplus_{p=\mathbf1,\sigma,\epsilon}
\mathcal V_{h_p}
\otimes\overline{\mathcal V}_{h_p}, \qquad (h_{\mathbf1},h_\sigma,h_\epsilon)
=
\left(0,\frac1{16},\frac12\right).
\end{equation}
Here $\mathcal V_h$ denotes the Virasoro highest-weight module of weight $h$, with null states quotiented out. The fusion rules and non-trivial primary OPE coefficient are
\begin{equation}
\sigma\times\sigma=\id+\epsilon,
\qquad
\sigma\times\epsilon=\sigma,
\qquad
\epsilon\times\epsilon=\id,
\qquad
C_{\sigma\sigma\epsilon}=\frac12.
\end{equation}
On the sphere, spin correlators are given by
\begin{equation}
\left\langle
\prod_{i=1}^{2m}\sigma(z_i,\bar z_i)
\right\rangle^2
=
2^{-m}
\sum_{\substack{e_i=\pm1\\\sum_i e_i=0}}
\prod_{i<j}
z_{ij}^{e_i e_j/4}\,
\bar z_{ij}^{e_i e_j/4}.
\end{equation}
with positive square root understood, and energy correlators are given by
\begin{equation}
\left\langle
\prod_{i=1}^{2m}\epsilon(z_i,\bar z_i)
\right\rangle
=
\operatorname{Pf}\!\left[\frac1{z_{ij}}\right]
\operatorname{Pf}\!\left[\frac1{\bar z_{ij}}\right].
\end{equation}
General mixed spin-energy correlators can be evaluated using the bosonization formula. Explicitly they are given by
\begin{equation}
\begin{aligned}
&\left\langle
\prod_{i=1}^{2m}\sigma(z_i,\bar z_i)
\prod_{r=1}^{p}\epsilon(w_r,\bar w_r)
\right\rangle^{\!2}
=
2^{-m}
\sum_{\substack{e_i=\pm1\\\sum_i e_i=0}}
\left[
\prod_{i<j}
z_{ij}^{e_ie_j/4}\bar z_{ij}^{e_ie_j/4}
\right]
\mathcal W_{\boldsymbol e}(\boldsymbol w;\boldsymbol z)
\mathcal W_{\boldsymbol e}(\bar{\boldsymbol w};\bar{\boldsymbol z}),
\end{aligned}
\end{equation}
with $\mathcal W_{\boldsymbol e}$ a Wick-contraction factor involving current insertions in the bosonization. The one- and two-energy factors are
\begin{equation}
\begin{aligned}
\mathcal W_{\boldsymbol e}(w;\boldsymbol z)
&=\frac12\sum_i\frac{e_i}{w-z_i},
\\
\mathcal W_{\boldsymbol e}(w_1,w_2;\boldsymbol z)
&=
\mathcal W_{\boldsymbol e}(w_1;\boldsymbol z)
\mathcal W_{\boldsymbol e}(w_2;\boldsymbol z)
+\frac1{(w_1-w_2)^2}.
\end{aligned}
\end{equation}
Correlators involving descendants are fixed by Virasoro Ward identities.

\paragraph{Modular integrals.} The modular-flowed pure-spin correlator is
\begin{equation}
\begin{aligned}
\left\langle
\prod_{j=0}^{2m-1}\sigma^{(is_j)}(0)
\right\rangle^{\!2}
&=
2^{-m}
\sum_{\substack{e_0,\ldots,e_{2m-1}=\pm1\\
\sum_j e_j=0}}
\prod_{0\le i<j\le2m-1}
\left[-i\sinh\frac{s_j-s_i}{2}\right]^{e_i e_j/2}
\\
&=
\prod_{0\le i<j\le2m-1}
\left[-i\sinh\frac{s_j-s_i}{2}\right]^{(-1)^{i+j}/2}.
\end{aligned}
\end{equation}
Under change of variable
\begin{equation}
y_j=\frac{e^{s_j}}{1+e^{s_j}},
\end{equation}
the resulting modular integral may be written as a Selberg-type integral with a rational insertion
\begin{equation}
\begin{aligned}
\mathcal I\bigl(\{(\sigma)^{2m}\}\bigr)
={}&
\frac{2^{m/2}e^{5\pi i(m-1)/4}}
{2m(2\pi)^{2m-2}}
\int_\Gamma
\prod_{j=1}^{2m-1}dy_j\,[y_j(1-y_j)]^{1/4}
\\
&\times
\prod_{1\le i<j\le2m-1}(y_j-y_i)^{1/2}
R_{2m}(\boldsymbol y),
\\
R_{2m}(\boldsymbol y)
={}&
\left[
\prod_{\substack{1\le i<j\le2m-1\\i+j\ {\rm odd}}}
(y_j-y_i)
\prod_{j=2}^{2m-1}(y_j-y_{j-1})
\right]^{-1}.
\end{aligned}
\label{eq:pure-spin-integral}
\end{equation}
Here the integration contour is
\begin{equation}
\begin{gathered}
\Gamma=\Gamma_1\times\cdots\times\Gamma_{2m-1},
\qquad
\Gamma_j:\quad
y_j(t)=\frac{e^{t+i\theta_j}}{1+e^{t+i\theta_j}},
\quad -\infty<t<\infty,
\\
-\pi<\theta_1<\cdots<\theta_{2m-1}<\pi,
\end{gathered}
\label{eq:y-contour}
\end{equation}
with each arc oriented from $0$ to $1$. Mixed correlators take similar form as~\cref{eq:pure-spin-integral} with different rational insertions. For example, with one energy insertion, the modular integral reads
\begin{equation}
\begin{aligned}
\mathcal I\bigl(\{(\sigma)^{2m},\epsilon\}\bigr)
={}&
\frac{2^{m/2}e^{i\pi(5m-2)/4}}
{(2\pi)^{2m-1}}
\int_\Gamma
\prod_{j=1}^{2m}dy_j\,[y_j(1-y_j)]^{1/4}
\\
&\times
\prod_{1\le i<j\le2m}(y_j-y_i)^{1/2}
R_{2m,1}(\boldsymbol y),
\\
R_{2m,1}(\boldsymbol y)
={}&
\frac{
\displaystyle\left[\sum_{j=1}^{2m}(-1)^j y_j\right]^2
}{
\displaystyle
\prod_{\substack{1\le i<j\le2m\\i+j\ {\rm odd}}}(y_j-y_i)
\prod_{j=2}^{2m}(y_j-y_{j-1})
}.
\end{aligned}
\label{eq:mixed-one-energy-integral}
\end{equation}
The contour is defined similarly to~\cref{eq:y-contour}. We have not obtained closed-form evaluations of the modular integrals such as~\cref{eq:pure-spin-integral,eq:mixed-one-energy-integral} for general $m$. Their $m=1$ specializations are given by the general two- and three-point results~\cref{eq:two-pt-sector-primary,eq: three-pt-sector-primary-result}.

\paragraph{Analytic reductions.} Some modular integrals admit analytic reductions that simplify subsequent numerical evaluations. For example,
\begin{equation}
\begin{aligned}
25\,\mathcal I\bigl(\{(\sigma)^3,\sigma_1\}\bigr)
+6\,\mathcal I\bigl(\{(\sigma)^4\}\bigr)
&=
-\frac{21}{32\pi^2}
\int_{\mathbb R^2}
\frac{du\,dv}
{\cosh^{5/2}(u/2)\cosh^{5/2}(v/2)}
=
-\frac{42}{\pi^2}
\mathcal I\bigl(\{(\sigma)^2\}\bigr)^2.
\end{aligned}
\end{equation}
This follows from writing the combined integrand in LHS as a total derivative; the remaining boundary term contribution factorizes into product of two-point primary contributions. Similar other analytic reductions are listed in~\cref{tab:ising-blocks}.

\paragraph{Numerical evaluations.} We evaluate the higher-dimensional modular integrals using randomized quasi-Monte Carlo integration with scrambled Sobol sequences. Independent randomizations provide estimates of the numerical uncertainty. The numerical values for $\mathcal I(X)$ up to radial order $\xi^{5/2}$ are listed in~\cref{tab:ising-blocks}.

\begin{table}[!tbp]
\centering
\footnotesize
\setlength{\tabcolsep}{4pt}
\renewcommand{\arraystretch}{1.1}
\begin{tabular}{@{}clcl@{}}
\hline
$\Delta_X$ & $X$ & $n=|X|$ & $\mathcal I(X)$\\
\hline
$1/4$ & $\{(\sigma)^2\}$ & $2$ &
$\frac{\sqrt\pi\,\Gamma(\tfrac54)}{4\Gamma(\tfrac74)}$\\
\hline
$1/2$ & $\{(\sigma)^4\}$ & $4$ & $0.0961290$\\
\hline
$3/4$ & $\{(\sigma)^6\}$ & $6$ & $0.0574440$\\
\hline
$1$ & $\{(\sigma)^8\}$ & $8$ & $0.049461(11)$\\
\hline
$5/4$ & $\{(\sigma)^{10}\}$ & $10$ & $0.052155(44)$\\
 & $\{\sigma,\sigma_1\}$ & $2$ &
$-\frac17\mathcal I(\{(\sigma)^2\})$\\
 & $\{(\sigma)^2,\epsilon\}$ & $3$ &
$-\frac1{14}\mathcal I(\{(\sigma)^2\})$\\
 & \multicolumn{2}{r}{coefficient of $\xi^{5/4}$} & $-0.041490(44)$\\
\hline
$3/2$ & $\{(\sigma)^{12}\}$ & $12$ & $0.062943(51)$\\
 & $\{(\sigma)^3,\sigma_1\}$ & $4$ &
$-\frac6{25}\mathcal I(\{(\sigma)^4\})
-\frac{42}{25\pi^2}\mathcal I(\{(\sigma)^2\})^2$\\
 & $\{(\sigma)^4,\epsilon\}$ & $5$ &
$-\frac7{25}\mathcal I(\{(\sigma)^4\})
-\frac{24}{25\pi^2}\mathcal I(\{(\sigma)^2\})^2$\\
 & \multicolumn{2}{r}{coefficient of $\xi^{3/2}$} & $-0.038129(51)$\\
\hline
$7/4$ & $\{(\sigma)^{14}\}$ & $14$ & $0.08370(16)$\\
 & $\{(\sigma)^5,\sigma_1\}$ & $6$ & $-0.0654056(71)$\\
 & $\{(\sigma)^6,\epsilon\}$ & $7$ & $-0.069386(64)$\\
 & \multicolumn{2}{r}{coefficient of $\xi^{7/4}$} & $-0.05109(17)$\\
\hline
$2$ & $\{(\sigma)^{16}\}$ & $16$ & $0.12091(49)$\\
 & $\{(\sigma)^7,\sigma_1\}$ & $8$ & $-0.089290(60)$\\
 & $\{(\sigma)^8,\epsilon\}$ & $9$ & $-0.11232(24)$\\
 & $\{(\epsilon)^2\}$ & $2$ & $\frac4{15}$\\
 & \multicolumn{2}{r}{coefficient of $\xi^{2}$} & $0.18597(55)$\\
\hline
$9/4$ & $\{(\sigma)^{18}\}$ & $18$ & $0.1883(21)$\\
 & $\{(\sigma)^9,\sigma_1\}$ & $10$ & $-0.13459(13)$\\
 & $\{(\sigma)^{10},\epsilon\}$ & $11$ & $-0.1946(24)$\\
 & $\{\sigma,\sigma_2\}$ & $2$ &
$\frac{15}{308}\mathcal I(\{(\sigma)^2\})$\\
 & $\{(\sigma_1)^2\}$ & $2$ &
$\frac{49}{44}\mathcal I(\{(\sigma)^2\})$\\
 & $\{\sigma,\sigma_1,\epsilon\}$ & $3$ &
$-\frac{107}{154}\mathcal I(\{(\sigma)^2\})$\\
 & $\{(\sigma)^2,\epsilon_1\}$ & $3$ &
$\frac5{308}\mathcal I(\{(\sigma)^2\})$\\
 & $\{(\sigma)^2,(\epsilon)^2\}$ & $4$ &
$\frac{617}{1232}\mathcal I(\{(\sigma)^2\})$\\
 & $\{(\sigma)^2,\mathbf1_2\}$ & $3$ &
$\frac9{1232}\mathcal I(\{(\sigma)^2\})$\\
 & \multicolumn{2}{r}{coefficient of $\xi^{9/4}$} & $0.2926(32)$\\
\hline
$5/2$ & $\{(\sigma)^{20}\}$ & $20$ & $0.2979(89)$\\
 & $\{(\sigma)^{11},\sigma_1\}$ & $12$ & $-0.21893(47)$\\
 & $\{(\sigma)^{12},\epsilon\}$ & $13$ & $-0.34085(265)$\\
 & $\{(\sigma)^3,\sigma_2\}+\{(\sigma)^2,(\sigma_1)^2\}$ & $4$ & $0.323748(4)$\\
 & $\{(\sigma)^3,\sigma_1,\epsilon\}+\{(\sigma)^4,\epsilon_1\}$ & $5$ & $-0.34532(7)$\\
 & $\{(\sigma)^4,(\epsilon)^2\}$ & $6$ & $0.32626(2)$\\
 & $\{(\sigma)^4,\mathbf1_2\}$ & $5$ & $0.005097(4)$\\
 & \multicolumn{2}{r}{coefficient of $\xi^{5/2}$} & $0.0479(93)$\\
\hline
\end{tabular}
\caption{Modular-integral coefficients $\mathcal I(X)$ for Ising \CFT{2}.
An entry $X_1+X_2$ gives value for $\mathcal I(X_1)+\mathcal I(X_2)$.}
\label{tab:ising-blocks}
\begingroup
\renewcommand{\showlabelsetlabel}[1]{}
\label{tab:ising-blocks-low}
\label{tab:ising-blocks-high}
\endgroup
\end{table}

\begin{figure}[tbp]
\centering
\begin{subfigure}{\linewidth}
\includegraphics[width=\linewidth]{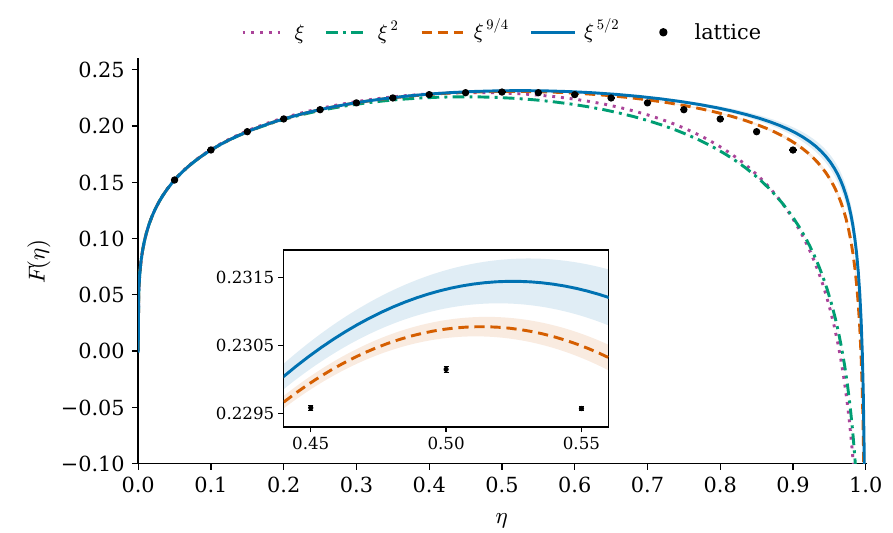}
\caption{Comparison between truncations of~\cref{eq:ising-mi-expansion} and lattice data. The inset shows a magnified comparison near $\eta=1/2$ for the two highest-order truncations. The negative values near $\eta =1$ is an unphysical artifact due to truncation.}\label{fig:ising-mi-comparison}
\end{subfigure}\par\vspace{4mm}
\begin{subfigure}{\linewidth}
\includegraphics[width=\linewidth]{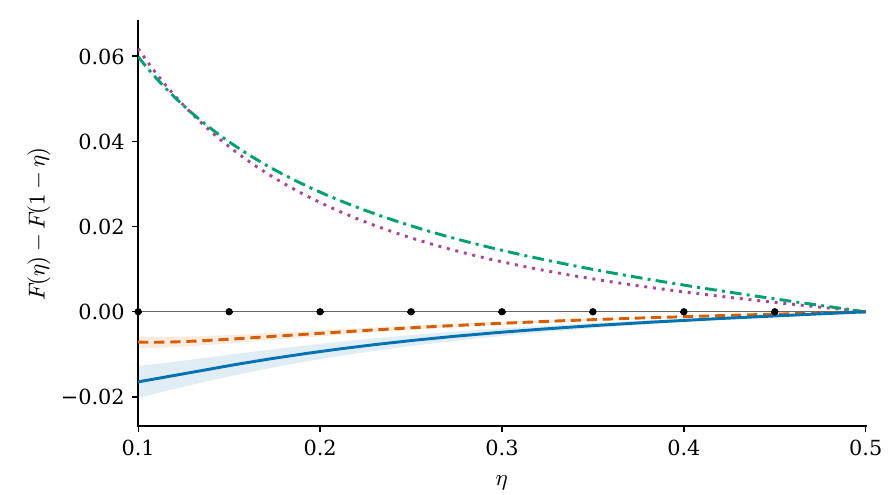}
\caption{Check of crossing symmetry~\cref{eq:mi-crossing-symmetry}.}\label{fig:ising-mi-crossing-test}
\end{subfigure}
\caption{The crossing-symmetric part $F(\eta)$ of mutual information
in Ising \CFT{2}. Shaded bands show uncertainties from numerical
evaluation of the modular integrals. Lattice points are obtained
by extrapolating spin chain results to continuum limit, with empirical uncertainty bars.}
\label{fig:ising-mi-crossing}
\end{figure}

\paragraph{Long distance expansion.} Assembling the modular integrals in~\cref{tab:ising-blocks}, we obtain long distance expansion of Ising mutual information as
\begin{equation}\label{eq:ising-mi-expansion}
\begin{aligned}
I(\xi)\simeq{}&
\frac{\sqrt\pi\,\Gamma(\tfrac54)}
{4\Gamma(\tfrac74)}\,\xi^{1/4}
+0.0961290\,\xi^{1/2}
+0.0574440\,\xi^{3/4}
\\
&+0.049461(11)\,\xi
-0.041490(44)\,\xi^{5/4}
-0.038129(51)\,\xi^{3/2}
\\
&-0.05109(17)\,\xi^{7/4}
+0.18597(55)\,\xi^2
+0.2926(32)\,\xi^{9/4}
+0.0479(93)\,\xi^{5/2} + \mathcal{O}(\xi^{11/4}).
\end{aligned}
\end{equation}
The comparison with lattice data is shown
in~\cref{fig:ising-mi-crossing}; the lattice data are obtained
using the method explained in~\cref{app:ising-lattice}.
The two highest-order truncations give good agreement with
lattice data at finite separation and substantially reduce
the violation of crossing symmetry~\cref{eq:mi-crossing-symmetry},
although the improvement with radial order is not monotonic.
These results indicate that the truncated expansion remains
useful beyond the strict long-distance regime, while its
convergence remains an open question.

\subsection{Free compact boson}\label{sec:other-CFT2}
The free compact boson has central charge $c=1$, with Hilbert space decomposed into sectors labeled by momentum $m$ and winding $w$:
\begin{equation}
\mathcal H_R
=
\bigoplus_{m,w\in\mathbb Z}
\mathcal F_p\otimes\overline{\mathcal F}_{\bar p},
\qquad
p=\frac{m/R+wR}{\sqrt2},
\qquad
\bar p=\frac{m/R-wR}{\sqrt2}.
\end{equation}
Here $\mathcal F_p$ denotes the Fock space generated by acting negative current modes $J_{-n}$ on a charge-$p$ highest-weight state of $U(1)$ current algebra. The currents $J$ and $\bar J$ belong to the neutral sector $m=w=0$. The corresponding vertex operator $V_{m,w}$ has dimensions
\begin{equation}
h_{m,w}=\frac{p^2}{2},
\qquad
\bar h_{m,w}=\frac{\bar p^2}{2}.
\end{equation}
The sphere correlators of vertex operators are given by
\begin{equation}
\left\langle
\prod_{i=1}^{k}V_{m_i,w_i}(z_i,\bar z_i)
\right\rangle
    =
\delta_{\sum_i m_i,0}\,
\delta_{\sum_i w_i,0}\,
\prod_{i<j}
z_{ij}^{p_i p_j}\,
\bar z_{ij}^{\bar p_i\bar p_j}.
\end{equation}
Here we don't keep track of cocycle phases, as they will cancel between the two correlator factors in~\cref{eq:mi-master}. The current two-point correlator is given by
\begin{equation}
\langle J(z)J(u)\rangle=\frac1{(z-u)^2},
\end{equation}
and similarly for anti-holomorphic counterparts. Higher-point current correlators follow from Wick contractions, while mixed current–vertex correlators are fixed by the current Ward identities. The operator words $X$ will consist of total-level blocks $V_{m,w;\ell}$ with dimension $\Delta_{m,w;\ell}
=
\frac{m^2}{2R^2}+\frac{w^2R^2}{2}+\ell$. For example, $V_{0,0;1}=\operatorname{span}\{J,\bar J\}$.

\paragraph{Self-dual radius.} Mutual information for free compact boson at self-dual radius $R=1$ is known in closed form from replica method~\cite{CCT09}
\begin{equation}
I_{\mathrm{sd}}(\eta)
=-\frac13\log(1-\eta)
=\frac43\sum_{r=0}^{\infty}\frac{\xi^{2r+1}}{2r+1},
\label{eq:sd-boson-mi}
\end{equation}
i.e., this is~\cref{eq:mi-crossing-function} with $c=1$ and vanishing $F(\eta)$. This also coincides with the mutual information of free massless Dirac fermion in NS vacuum obtained using continuum generalization of correlation matrix method~\cite{Casini:2009vk}. Using \cref{eq:mi-master}, we reproduce the expansion \cref{eq:sd-boson-mi} through order $\xi^4$, cf.~\cref{tab:boson-self-dual}. An all-orders proof should be possible, but we don't pursue it here.

\paragraph{Current contribution.} The contribution from $U(1)$ current is known exactly from continuum correlation matrix method~\cite{ACHP18}
\begin{equation}
I_J(\eta)
=-\frac13\log(1-\eta)
-i\pi\int_0^\infty d\omega\,
\frac{\omega}{\sinh^2(\pi\omega)}
\log\frac{
{}_2F_1(1+i\omega,-i\omega;1;\eta)
}{
{}_2F_1(1-i\omega,i\omega;1;\eta)
}.
\end{equation}
This subsector contribution doesn't obey crossing symmetry~\cref{eq:mi-crossing-symmetry}, and has divergent asymptotic long distance expansion~\cite{Abate:2025ywp}
\begin{equation}
I_J
\sim
\frac8{15}\xi^2
-\frac{32}{105}\xi^3
+\frac{16}{15}\xi^4
-\frac{288}{77}\xi^5
+\frac{1182088}{45045}\xi^6
-\frac{805696}{3003}\xi^7
+\cdots .
\label{eq:current-expansion-mi}
\end{equation}
Specializing to current contribution to~\cref{eq:mi-master}, we reproduce the expansion~\cref{eq:current-expansion-mi} through $\xi^7$, cf.~\cref{tab:boson-current}.

\paragraph{Generic radius.} At the self-dual radius, cancellations between current and vertex operator contributions produce the convergent expansion in \cref{eq:sd-boson-mi}. At generic radius, such cancellation is absent. The long-distance expansion thus generically contains the divergent current contribution \cref{eq:current-expansion-mi}.

\begin{table}[tbp]
\centering
\small
\begin{lrbox}{\leftcoefficienttable}
\begin{tabular}{@{}clcl@{}}
\hline
$\Delta_X$ & $X$ & $n=|X|$ & $\mathcal I(X)$\\
\hline
$1$ & $\{(\mathcal B_{1/2})^{2}\}$ & $2$ & $\frac{4}{3}$\\
\hline
$2$ & $\{\mathcal B_{1/2},\mathcal B_{3/2}\}$ & $2$ & $-\frac{8}{15}$\\
 & $\{(\mathcal B_{1})^{2}\}$ & $2$ & $\frac{8}{5}$\\
 & $\{(\mathcal B_{1/2})^{2},\mathcal B_{1}\}$ & $3$ & $-\frac{16}{5}$\\
 & $\{(\mathcal B_{1/2})^{4}\}$ & $4$ & $\frac{32}{15}$\\
 & \multicolumn{2}{r}{coefficient of $\xi^2$} & $0$\\
\hline
$3$ & $\{\mathcal B_{1/2},\mathcal B_{5/2}\}$ & $2$ & $\frac{8}{35}$\\
 & $\{\mathcal B_{1},\mathcal B_{2}\}$ & $2$ & $-\frac{32}{35}$\\
 & $\{(\mathcal B_{3/2})^{2}\}$ & $2$ & $\frac{172}{105}$\\
 & $\{(\mathcal B_{1/2})^{2},\mathcal B_{2}\}$ & $3$ & $\frac{32}{35}$\\
 & $\{\mathcal B_{1/2},\mathcal B_{1},\mathcal B_{3/2}\}$ & $3$ & $-\frac{32}{7}$\\
 & $\{(\mathcal B_{1})^{3}\}$ & $3$ & $-\frac{256}{105}$\\
 & $\{(\mathcal B_{1/2})^{3},\mathcal B_{3/2}\}$ & $4$ & $\frac{128}{105}$\\
 & $\{(\mathcal B_{1/2})^{2},(\mathcal B_{1})^{2}\}$ & $4$ & $\frac{864}{35}$\\
 & $\{(\mathcal B_{1/2})^{4},\mathcal B_{1}\}$ & $5$ & $-\frac{256}{7}$\\
 & $\{(\mathcal B_{1/2})^{6}\}$ & $6$ & $\frac{1024}{63}$\\
 & \multicolumn{2}{r}{coefficient of $\xi^3$} & $\frac{4}{9}$\\
\hline
$4$ & $\{\mathcal B_{1/2},\mathcal B_{7/2}\}$ & $2$ & $-\frac{8}{63}$\\
 & $\{\mathcal B_{1},\mathcal B_{3}\}$ & $2$ & $\frac{16}{35}$\\
 & $\{\mathcal B_{3/2},\mathcal B_{5/2}\}$ & $2$ & $-\frac{8}{7}$\\
 & $\{(\mathcal B_{2})^{2}\}$ & $2$ & $\frac{992}{315}$\\
 & $\{(\mathcal B_{1/2})^{2},\mathcal B_{3}\}$ & $3$ & $-\frac{16}{35}$\\
 & $\{\mathcal B_{1/2},\mathcal B_{1},\mathcal B_{5/2}\}$ & $3$ & $\frac{32}{21}$\\
 & $\{\mathcal B_{1/2},\mathcal B_{3/2},\mathcal B_{2}\}$ & $3$ & $-\frac{1984}{315}$\\
 & $\{(\mathcal B_{1})^{2},\mathcal B_{2}\}$ & $3$ & $-\frac{128}{35}$\\
 & $\{\mathcal B_{1},(\mathcal B_{3/2})^{2}\}$ & $3$ & $-\frac{16}{5}$\\
 & $\{(\mathcal B_{1/2})^{3},\mathcal B_{5/2}\}$ & $4$ & $-\frac{128}{315}$\\
 & $\{(\mathcal B_{1/2})^{2},\mathcal B_{1},\mathcal B_{2}\}$ & $4$ & $-\frac{128}{21}$\\
 & $\{(\mathcal B_{1/2})^{2},(\mathcal B_{3/2})^{2}\}$ & $4$ & $\frac{6848}{315}$\\
 & $\{\mathcal B_{1/2},(\mathcal B_{1})^{2},\mathcal B_{3/2}\}$ & $4$ & $\frac{5056}{105}$\\
 & $\{(\mathcal B_{1})^{4}\}$ & $4$ & $\frac{2176}{105}$\\
 & $\{(\mathcal B_{1/2})^{4},\mathcal B_{2}\}$ & $5$ & $\frac{1024}{105}$\\
 & $\{(\mathcal B_{1/2})^{3},\mathcal B_{1},\mathcal B_{3/2}\}$ & $5$ & $-\frac{2048}{15}$\\
 & $\{(\mathcal B_{1/2})^{2},(\mathcal B_{1})^{3}\}$ & $5$ & $-\frac{10112}{35}$\\
 & $\{(\mathcal B_{1/2})^{5},\mathcal B_{3/2}\}$ & $6$ & $\frac{2048}{35}$\\
 & $\{(\mathcal B_{1/2})^{4},(\mathcal B_{1})^{2}\}$ & $6$ & $\frac{17408}{21}$\\
 & $\{(\mathcal B_{1/2})^{6},\mathcal B_{1}\}$ & $7$ & $-\frac{4096}{5}$\\
 & $\{(\mathcal B_{1/2})^{8}\}$ & $8$ & $\frac{4096}{15}$\\
 & \multicolumn{2}{r}{coefficient of $\xi^4$} & $0$\\
\hline
\end{tabular}
\end{lrbox}
\begin{lrbox}{\rightcoefficienttable}
\begin{tabular}{@{}clcl@{}}
\hline
$\Delta_X$ & $X$ & $n=|X|$ & $\mathcal I(X)$\\
\hline
$2$ & $\{(\mathsf J_{1})^{2}\}$ & $2$ & $\frac{8}{15}$\\
\hline
$3$ & $\{\mathsf J_{1},\mathsf J_{2}\}$ & $2$ & $-\frac{32}{105}$\\
 & $\{(\mathsf J_{1})^{3}\}$ & $3$ & $0$\\
 & \multicolumn{2}{r}{coefficient of $\xi^3$} & $-\frac{32}{105}$\\
\hline
$4$ & $\{\mathsf J_{1},\mathsf J_{3}\}$ & $2$ & $\frac{16}{105}$\\
 & $\{(\mathsf J_{2})^{2}\}$ & $2$ & $\frac{32}{35}$\\
 & $\{(\mathsf J_{1})^{2},\mathsf J_{2}\}$ & $3$ & $-\frac{128}{105}$\\
 & $\{(\mathsf J_{1})^{4}\}$ & $4$ & $\frac{128}{105}$\\
 & \multicolumn{2}{r}{coefficient of $\xi^4$} & $\frac{16}{15}$\\
\hline
$5$ & $\{\mathsf J_{1},\mathsf J_{4}\}$ & $2$ & $-\frac{64}{693}$\\
 & $\{\mathsf J_{2},\mathsf J_{3}\}$ & $2$ & $-\frac{160}{231}$\\
 & $\{(\mathsf J_{1})^{2},\mathsf J_{3}\}$ & $3$ & $\frac{512}{1155}$\\
 & $\{\mathsf J_{1},(\mathsf J_{2})^{2}\}$ & $3$ & $\frac{512}{1155}$\\
 & $\{(\mathsf J_{1})^{3},\mathsf J_{2}\}$ & $4$ & $-\frac{13312}{3465}$\\
 & $\{(\mathsf J_{1})^{5}\}$ & $5$ & $0$\\
 & \multicolumn{2}{r}{coefficient of $\xi^5$} & $-\frac{288}{77}$\\
\hline
$6$ & $\{\mathsf J_{1},\mathsf J_{5}\}$ & $2$ & $\frac{80}{1287}$\\
 & $\{\mathsf J_{2},\mathsf J_{4}\}$ & $2$ & $\frac{128}{315}$\\
 & $\{(\mathsf J_{3})^{2}\}$ & $2$ & $\frac{7256}{5005}$\\
 & $\{(\mathsf J_{1})^{2},\mathsf J_{4}\}$ & $3$ & $-\frac{256}{1001}$\\
 & $\{\mathsf J_{1},\mathsf J_{2},\mathsf J_{3}\}$ & $3$ & $-\frac{80128}{15015}$\\
 & $\{(\mathsf J_{2})^{3}\}$ & $3$ & $-\frac{11776}{3003}$\\
 & $\{(\mathsf J_{1})^{3},\mathsf J_{3}\}$ & $4$ & $\frac{76288}{15015}$\\
 & $\{(\mathsf J_{1})^{2},(\mathsf J_{2})^{2}\}$ & $4$ & $\frac{1024}{21}$\\
 & $\{(\mathsf J_{1})^{4},\mathsf J_{2}\}$ & $5$ & $-\frac{176128}{3465}$\\
 & $\{(\mathsf J_{1})^{6}\}$ & $6$ & $\frac{462848}{15015}$\\
 & \multicolumn{2}{r}{coefficient of $\xi^6$} & $\frac{1182088}{45045}$\\
\hline
$7$ & $\{\mathsf J_{1},\mathsf J_{6}\}$ & $2$ & $-\frac{32}{715}$\\
 & $\{\mathsf J_{2},\mathsf J_{5}\}$ & $2$ & $-\frac{32}{117}$\\
 & $\{\mathsf J_{3},\mathsf J_{4}\}$ & $2$ & $-\frac{18496}{15015}$\\
 & $\{(\mathsf J_{1})^{2},\mathsf J_{5}\}$ & $3$ & $\frac{512}{3003}$\\
 & $\{\mathsf J_{1},\mathsf J_{2},\mathsf J_{4}\}$ & $3$ & $\frac{32768}{15015}$\\
 & $\{\mathsf J_{1},(\mathsf J_{3})^{2}\}$ & $3$ & $\frac{1024}{715}$\\
 & $\{(\mathsf J_{2})^{2},\mathsf J_{3}\}$ & $3$ & $\frac{14848}{5005}$\\
 & $\{(\mathsf J_{1})^{3},\mathsf J_{4}\}$ & $4$ & $-\frac{14336}{6435}$\\
 & $\{(\mathsf J_{1})^{2},\mathsf J_{2},\mathsf J_{3}\}$ & $4$ & $-\frac{1076224}{15015}$\\
 & $\{\mathsf J_{1},(\mathsf J_{2})^{3}\}$ & $4$ & $-\frac{1675264}{15015}$\\
 & $\{(\mathsf J_{1})^{4},\mathsf J_{3}\}$ & $5$ & $\frac{475136}{15015}$\\
 & $\{(\mathsf J_{1})^{3},(\mathsf J_{2})^{2}\}$ & $5$ & $\frac{7143424}{45045}$\\
 & $\{(\mathsf J_{1})^{5},\mathsf J_{2}\}$ & $6$ & $-\frac{278528}{1001}$\\
 & $\{(\mathsf J_{1})^{7}\}$ & $7$ & $0$\\
 & \multicolumn{2}{r}{coefficient of $\xi^7$} & $-\frac{805696}{3003}$\\
\hline
\end{tabular}
\end{lrbox}
\setcoefficientpanelheight
\begin{subtable}[t]{0.49\textwidth}
\centering
\vspace{0pt}
\coefficientpanelbody{\leftcoefficienttable}
\caption{Self-dual radius}
\label{tab:boson-self-dual}
\end{subtable}\hfill%
\begin{subtable}[t]{0.49\textwidth}
\centering
\vspace{0pt}
\coefficientpanelbody{\rightcoefficienttable}
\caption{Current sector}
\label{tab:boson-current}
\end{subtable}
\caption{Modular-integral coefficients $\mathcal I(X)$ for the free
compact boson \CFT{2}: (a) the full theory at the self-dual radius;
(b) the current sector. The operator words are built from the complete
dimension blocks
$\mathcal B_\Delta=\bigoplus_{(m^2+w^2)/2+\ell=\Delta}V_{m,w;\ell}$ in (a), and from the current blocks
$\mathsf J_\ell=V_{0,0;\ell}$ in (b).}
\label{tab:sd-boson-dimension-blocks}
\end{table}

 \subsection{Relation with replica method}\label{sec:replica-relation}
 In replica method, one computes Renyi entropy $\operatorname{Tr} \rho^N_{AB}$, and obtains mutual information by analytic continuation in replica number $N$. We denote the normalized Renyi moment by
 \begin{equation}
\mathcal R_N(A:B)
=
\frac{\operatorname{Tr}\rho_{AB}^{N}}
{\operatorname{Tr}\rho_A^{N}\operatorname{Tr}\rho_B^{N}}.
 \end{equation}
The mutual information is obtained by continuation
\begin{equation}
I(A:B)
=
\left.\partial_N\mathcal R_N(A:B)\right|_{N=1}.
\end{equation}
In \CFT{2}, the Renyi entropies are computed by twist operator correlators in cyclic orbifold $\mathrm{CFT}^{\otimes N}/\mathbb Z_N$~\cite{Calabrese:2004eu,CCT09}. In concentric frame this is given by
\begin{equation}
\mathcal R_N(\xi)
=
\frac{
\left\langle
\widetilde{\mathcal T}_N(-1)\,
\mathcal T_N(-\xi)\,
\widetilde{\mathcal T}_N(\xi)\,
\mathcal T_N(1)
\right\rangle
}{
\left\langle
\mathcal T_N(-\xi)\widetilde{\mathcal T}_N(\xi)
\right\rangle
\left\langle
\widetilde{\mathcal T}_N(-1)\mathcal T_N(1)
\right\rangle
}.
\label{eq:replica-twist-ratio}
\end{equation}
Here $\mathcal T_N$ and $\widetilde{\mathcal T}_N$ are twist and anti-twist operators; they are primary scalar operators with dimension $h_N = \bar{h}_N= \frac{c}{24}(N-N^{-1})$. The two-point function of twist operators is fixed by conformal symmetry, and gives the universal single interval entropy in \CFT{2}~\cite{Calabrese:2004eu}. The non-trivial part of~\cref{eq:replica-twist-ratio} is therefore the twist-operator four-point function. From covering space perspective, twist operator four-point function can be viewed as genus $g=N-1$ partition function of \CFT{2}. Obtaining mutual information at finite $\xi$ by continuation of~\cref{eq:replica-twist-ratio} therefore requires analytic continuation of \CFT{2} partition function in genus $g$; this challenge remains generally unresolved.

More progress can be made by considering long distance expansion of~\cref{eq:replica-twist-ratio}, which can be systematically computed at fixed $N$ from twist operator OPE or expanding genus $g=N-1$ partition function. The expansion generally takes the form
\begin{equation}
\mathcal R_N(\xi)
\sim
1+
\sum_{\substack{1\le |X|\le N}}
\xi^{\Delta_X}\mathcal C_N(X).
\label{eq:renyi-expansion}
\end{equation}
Here we use the same operator word notation as in~\cref{eq:mi-X-sum}. The expansion coefficient $\mathcal C_N(X)$ is sum of OPE coefficients associated with operators of the form
\begin{eqnarray}
\mathcal O_X\sim
\mathcal O_{a_1}\otimes\cdots\otimes\mathcal O_{a_n}
\otimes\id^{\otimes(N-n)}
+\text{cyclic images},
\qquad
a_j\in X_j, \quad \abs{X}=n,
\end{eqnarray}
appearing in twist operator OPE. The OPE coefficients can be evaluated using covering space method as~\cite{Calabrese:2010he}
\begin{equation}
\begin{aligned}
\mathcal C_N(X)
={}&
\frac Nn
\sum_{\boldsymbol\lambda\in\operatorname{Perm}(X)}
\sum_{\substack{a_k,b_k\in\lambda_k\\k=0,\ldots,n-1}}
\left[\prod_{k=0}^{n-1}(G^{-1})_{b_ka_k}\right]
\sum_{0<j_1<\cdots<j_{n-1}<N}
\\
&\times
\left\langle
\prod_{k=0}^{n-1}
\bigl[f_N\cdot\mathcal O_{a_k}^{\dagger}\bigr]^{
(\pi-2\pi j_k/N)}(0)
\right\rangle
\left\langle
\prod_{k=0}^{n-1}
\bigl[f_N\cdot\mathcal O_{b_k}\bigr]^{
(2\pi j_k/N)}(0)
\right\rangle,
\qquad j_0=0.
\end{aligned}
\label{eq:CN-covering-space}
\end{equation}
Here $f_N$ is the unit interval covering map
\begin{equation}
f_N(z)
=
\frac{
\left(\dfrac{1+z}{1-z}\right)^{1/N}-1
}{
\left(\dfrac{1+z}{1-z}\right)^{1/N}+1
},
\end{equation}
superscript denotes the modular rotation as in~\cref{eq:flowed-field}, and $f_N\cdot\mathcal O$ denotes the action of local conformal transformation $f_N$. Here modular rotation appears because rotation by modular angle $2\pi/N$ in uniformizing coordinate changes between sheets. For example, for $X = \{\mathcal{O},\mathcal{O}\}$ with scalar primary $\mathcal{O}$ of dimension $\Delta$, the OPE coefficient is
\begin{equation}
\label{eq:CN-OPE-two-primary}
\begin{aligned}
\mathcal C_N\bigl(\{(\mathcal O)^2\}\bigr)
={}&
\frac{N}{2N^{4\Delta}}
\sum_{j=1}^{N-1}
\left\langle
\mathcal O^{(\pi)}(0)
\mathcal O^{(\pi-2\pi j/N)}(0)
\right\rangle
\left\langle
\mathcal O(0)
\mathcal O^{(2\pi j/N)}(0)
\right\rangle
\\
={}&
\frac{N}{2N^{4\Delta}}
\sum_{j=1}^{N-1}
\left[\sin\frac{\pi j}{N}\right]^{-4\Delta}.
\end{aligned}
\end{equation}

The long distance expansion of mutual information is then obtained from termwise continuation of~\cref{eq:renyi-expansion}
\begin{equation}
  I(\xi) \sim \sum_X \xi^{\Delta_X}\mathcal C^\prime_1(X),
\end{equation}
where $\mathcal{C}^\prime_1(X) = \left.\partial_N\mathcal C_N(X)\right|_{N=1}$ denotes the formal continuation. Comparing with~\cref{eq:mi-X-sum}, we see that our modular integral $\mathcal{I}(X)$ precisely provides the continuation of twist operator OPE coefficient:
\begin{equation}
  \boxed{
    \mathcal{C}^\prime_1(X) = \mathcal{I}(X).
  }
\label{eq:C-equals-I}
\end{equation}
Moreover, the response order $n=\abs{X}$ in~\cref{eq:mi-master} corresponds to, in replica method, having $n$ of $N$ copies of CFT with operator insertions before the continuation in $N$. For $X = \{\mathcal{O},\mathcal{O}\}$, the identification~\cref{eq:C-equals-I} is derived in~\cite{Calabrese:2010he} by directly studying analytic continuation of~\cref{eq:CN-OPE-two-primary}. For generic operator word $X$, we are not aware of a direct derivation of~\cref{eq:C-equals-I} from analytic continuation of $\mathcal{C}_{N}(X)$. Such a derivation may be possible by generalizing the argument in~\cite{Agon:2015ftl}.

In practice, $\mathcal{C}^\prime_1(X)$ for generic $X$ can be obtained using numerical continuation~\cite{DeNobili:2015dla,RTC18}. In general, for $\abs{X}=n$, the OPE coefficient and its formal continuation takes the form
\begin{equation}
\mathcal C_N(X)
=
N(N-1)\cdots(N-n+1)\,f_X(N),
\qquad
\mathcal C'_1(X)
=
(-1)^{n-2}(n-2)!\,f_X(1).
\end{equation}
Numerical continuation therefore amounts to extrapolating $f_X(1)$ from its values at integers $N\ge n$. In the following we compare numerically extrapolated $\mathcal C'_1(X)$ with our modular integral $\mathcal I(X)$ in the Ising example.

\paragraph{Ising.} The normalized Renyi moment for Ising is known in closed form~\cite{Calabrese:2010he}
\begin{equation}
\mathcal R_N(\xi)
=
\left(\frac{1+\xi}{1-\xi}\right)^{(N-N^{-1})/6}
\frac{
\displaystyle
\sum_{\boldsymbol\varepsilon,\boldsymbol\delta
\in\{0,\frac12\}^{N-1}}
\left|
\Theta\!\begin{bmatrix}
\boldsymbol\varepsilon\\
\boldsymbol\delta
\end{bmatrix}
(\boldsymbol0\mid\tau_N)
\right|
}{
2^{N-1}\Theta(\boldsymbol0\mid\tau_N)
}.
\label{eq:Ising-renyi}
\end{equation}
Here $\Theta$ is the genus-$(N-1)$ Riemann theta function, the denominator has zero characteristic, and the period matrix is
\begin{equation}
(\tau_N)_{rs}
=
\frac{2i}{N}
\sum_{k=1}^{N-1}
\sin\frac{\pi k}{N}\,
\cos\frac{2\pi k(r-s)}{N}\,
\frac{
{}_2F_1(k/N,1-k/N;1;1-\eta)
}{
{}_2F_1(k/N,1-k/N;1;\eta)
},
\quad
1\le r,s\le N-1.
\end{equation}
Its long distance expansion is given by
\begin{equation}
\begin{aligned}
\mathcal R_N(\xi)
={}&1+
\sum_{m=1}^{4}
\xi^{m/4}\,
\mathcal C_N\bigl(\{(\sigma)^{2m}\}\bigr)
\\
&+\xi^{5/4}
\Big[
\mathcal C_N\bigl(\{(\sigma)^{10}\}\bigr)
+\mathcal C_N\bigl(\{\sigma,\sigma_1\}\bigr)
+\mathcal C_N\bigl(\{(\sigma)^2,\epsilon\}\bigr)
\Big]
+O(\xi^{3/2}),
\end{aligned}
\end{equation}
where $\sigma_1$ includes contribution from both $\partial \sigma$ and $\bar{\partial}\sigma$, and coefficients with $|X|>N$ vanish. The OPE coefficients are given explicitly by
\begin{equation}
\begin{aligned}
\mathcal C_N\bigl(\{(\sigma)^{2m}\}\bigr)
={}&
\frac{N^{1-m/2}}{2m}
\sum_{0=j_0<j_1<\cdots<j_{2m-1}<N}
\prod_{0\le k<l\le2m-1}
\left[
\sin\frac{\pi(j_l-j_k)}N
\right]^{(-1)^{k+l}/2},
\\[2mm]
\mathcal C_N\bigl(\{\sigma,\sigma_1\}\bigr)
={}&
\frac1{4N^{3/2}}
\sum_{j=1}^{N-1}
\frac{\cos^2(\pi j/N)}
{\sin^{5/2}(\pi j/N)},
\\[2mm]
\mathcal C_N\bigl(\{(\sigma)^2,\epsilon\}\bigr)
={}&
\frac1{4N^{3/2}}
\sum_{0<j<k<N}
\frac{\sin^{3/2}(\pi j/N)}
{\sin^2(\pi k/N)\sin^2(\pi(k-j)/N)}.
\end{aligned}
\end{equation}
These can be obtained by either directly expanding~\cref{eq:Ising-renyi} or using the covering space formula~\cref{eq:CN-covering-space}. Comparison between numerically continued $\mathcal{C}^\prime_1(X)$ and modular integral $\mathcal{I}(X)$ is given in~\cref{tab:ising-replica-comparison}, and we find good agreement.

\begin{table}[tbp]
\centering
\renewcommand{\arraystretch}{1.4}
\begin{tabular}{@{}ccc@{}}
\hline
\noalign{\vskip 4pt}
$X$ & \shortstack{$\mathcal C'_1(X)$}
& \shortstack{$\mathcal I(X)$}\\
\hline
$\{(\sigma)^4\}$ & $0.09613$ & $0.0961290$\\
$\{(\sigma)^6\}$ & $0.05744$ & $0.0574440$\\
$\{\sigma,\sigma_1\}$ & $-0.06243$ &
$\displaystyle -\frac{\sqrt\pi\,\Gamma(\tfrac54)}{28\Gamma(\tfrac74)}
\simeq -0.0624299$\\[4pt]
\hline
\end{tabular}
\caption{Comparison between numerically continued $\mathcal{C}^\prime_1(X)$ and modular integral $\mathcal{I}(X)$. We extrapolate the zero-factored
coefficients $f_X(N)$ from integer replica numbers $N\le36$ to $N=1$
using rational fits.}
\label{tab:ising-replica-comparison}
\end{table}
\FloatBarrier

\paragraph{Higher dimensions.} The identification~\cref{eq:C-equals-I} also applies in $d>2$, where $\mathcal C_N(X)$ are OPE coefficients in the higher-dimensional twist operator OPE. They involve thermal
correlators on hyperbolic space, which at generic $N$ cannot
be obtained from flat-space vacuum correlators by a conformal
map, and are generally not known in closed form.
For the scalar primary two-point contribution, the continuation
was derived in~\cite{Agon:2015ftl}.

\section{Free scalar \CFT{d}}\label{sec:scalar-CFTd}
In this section we apply~\cref{eq:mi-master} to the free scalar
\CFT{d} with $d\geq3$, providing an analytical check of our
construction beyond $d=2$. Long-distance coefficients
have been obtained using replica methods~\cite{Agon:2015twa,Chen:2017hbk}
and a systematic replica-free method exploiting the Gaussian
structure of the theory~\cite{Buchanan:2024tlm}.

\subsection{Operator words and modular integrals}
The free scalar \CFT{d} has local operators generated from scalar $\phi$ with dimension
\begin{equation}
  \Delta_\phi = \frac{d-2}{2}
\end{equation}
and null state condition
\begin{equation}
\Box\phi=0.
\end{equation}
Local operators are normal-ordered products of derivatives of $\phi$, and their correlation functions are determined by Wick’s theorem. To organize the expansion, we group the operators by their field number $N$ and total derivative level $L$:
\begin{equation}
\Phi_{N,L}
:=
\operatorname{span}
\left\{
:\prod_{i=1}^{N}\partial^{\ell_i}\phi:
\ \middle|\
\sum_{i=1}^{N}\ell_i=L
\right\},
\qquad
\Delta_{N,L}=N\frac{d-2}{2}+L.
\end{equation}
Here $\partial^\ell\phi$ denotes an $\ell$-fold derivative with tensor indices suppressed;  equation-of-motion null states are removed. Different $(N,L)$ sectors are mutually orthogonal Gram block. We abbreviate $\Phi_{N,0}$ as $\Phi_N$, with $\Phi_1=\phi$. For example, the $(N,L)=(1,2)$ block has one null state removed, with dimension $\dim\Phi_{1,2}=d(d+1)/2-1$.
The operators words are then organized in terms of the Gram bocks $\Phi_{N,L}$. The modular-integral coefficients can be evaluated analytically, and are given in~\cref{tab:free-scalar-blocks} for $d=3,4$.

\begin{table}[tbp]
\centering
\small
\renewcommand{\arraystretch}{1.3}
\begin{lrbox}{\leftcoefficienttable}
\begin{tabular}{@{}clcl@{}}
\hline
$\Delta_X$ & $X$ & $n=|X|$ & $\mathcal I(X)$\\
\hline
$1$ & $\{(\phi)^2\}$ & $2$ & $\frac13$\\
\hline
$2$ & $\{\phi,\Phi_{1,1}\}$ & $2$ & $-\frac2{15}$\\
 & $\{(\Phi_2)^2\}$ & $2$ & $\frac4{15}$\\
 & $\{(\phi)^2,\Phi_2\}$ & $3$ & $-\frac8{15}$\\
 & $\{(\phi)^4\}$ & $4$ & $\frac{14}{45}+\frac4{3\pi^2}$\\
 & \multicolumn{2}{r}{coefficient of $\xi^2$} & $-\frac4{45}+\frac4{3\pi^2}$\\
\hline
$3$ & $\{\phi,\Phi_{1,2}\}$ & $2$ & $\frac2{35}$\\
 & $\{(\Phi_{1,1})^2\}$ & $2$ & $\frac{67}{105}$\\
 & $\{\Phi_2,\Phi_{2,1}\}$ & $2$ & $-\frac{16}{105}$\\
 & $\{(\Phi_3)^2\}$ & $2$ & $\frac8{35}$\\
 & $\{\phi,\Phi_{1,1},\Phi_2\}$ & $3$ & $\frac{16}{105}$\\
 & $\{(\phi)^2,\Phi_{2,1}\}$ & $3$ & $\frac{16}{105}$\\
 & $\{(\Phi_2)^3\}$ & $3$ & $-\frac{256}{315}$\\
 & $\{\phi,\Phi_2,\Phi_3\}$ & $3$ & $-\frac{48}{35}$\\
 & $\{(\phi)^3,\Phi_{1,1}\}$ & $4$ & $-\frac{88}{315}-\frac8{3\pi^2}$\\
 & $\{(\phi)^2,(\Phi_2)^2\}$ & $4$ & $\frac{536}{105}+\frac{64}{3\pi^2}$\\
 & $\{(\phi)^3,\Phi_3\}$ & $4$ & $\frac{32}{35}$\\
 & $\{(\phi)^4,\Phi_2\}$ & $5$ & $-\frac{32}{5}-\frac{128}{3\pi^2}$\\
 & $\{(\phi)^6\}$ & $6$ & $\begin{array}[t]{@{}l@{}}\frac{752}{315}+\frac{208}{9\pi^2}\\{}+\frac{16}{3\pi^4}\end{array}$\\
 & \multicolumn{2}{r}{coefficient of $\xi^3$} & $\begin{array}[t]{@{}l@{}}\frac{13}{21}-\frac8{9\pi^2}\\{}+\frac{16}{3\pi^4}\end{array}$\\
\hline
\end{tabular}
\end{lrbox}
\begin{lrbox}{\rightcoefficienttable}
\begin{tabular}{@{}clcl@{}}
\hline
$\Delta_X$ & $X$ & $n=|X|$ & $\mathcal I(X)$\\
\hline
$2$ & $\{(\phi)^2\}$ & $2$ & $\frac4{15}$\\
\hline
$3$ & $\{\phi,\Phi_{1,1}\}$ & $2$ & $-\frac{16}{105}$\\
\hline
$4$ & $\{\phi,\Phi_{1,2}\}$ & $2$ & $\frac8{105}$\\
 & $\{(\Phi_{1,1})^2\}$ & $2$ & $\frac{16}{21}$\\
 & $\{(\Phi_2)^2\}$ & $2$ & $\frac{64}{315}$\\
 & $\{(\phi)^2,\Phi_2\}$ & $3$ & $-\frac{128}{315}$\\
 & $\{(\phi)^4\}$ & $4$ & $\frac{32}{63}$\\
 & \multicolumn{2}{r}{coefficient of $\xi^4$} & $\frac87$\\
\hline
$5$ & $\{\phi,\Phi_{1,3}\}$ & $2$ & $-\frac{32}{693}$\\
 & $\{\Phi_{1,1},\Phi_{1,2}\}$ & $2$ & $-\frac{656}{1155}$\\
 & $\{\Phi_2,\Phi_{2,1}\}$ & $2$ & $-\frac{512}{3465}$\\
 & $\{(\phi)^2,\Phi_{2,1}\}$ & $3$ & $\frac{512}{3465}$\\
 & $\{\phi,\Phi_{1,1},\Phi_2\}$ & $3$ & $\frac{512}{3465}$\\
 & $\{(\phi)^3,\Phi_{1,1}\}$ & $4$ & $-\frac{1280}{693}$\\
 & \multicolumn{2}{r}{coefficient of $\xi^5$} & $-\frac{2672}{1155}$\\
\hline
\end{tabular}
\end{lrbox}
\setcoefficientpanelheight
\begin{subtable}[t]{0.56\textwidth}
\centering
\vspace{0pt}
\coefficientpanelbody{\leftcoefficienttable}
\caption{$d=3$}
\label{tab:free-scalar-d3}
\end{subtable}\hfill
\begin{subtable}[t]{0.42\textwidth}
\centering
\vspace{0pt}
\coefficientpanelbody{\rightcoefficienttable}
\caption{$d=4$}
\label{tab:free-scalar-d4}
\end{subtable}
\caption{Modular-integral coefficients $\mathcal I(X)$ for free
scalar \CFT{d}.}
\label{tab:free-scalar-blocks}
\end{table}

\subsection{Long distance expansion}
The long distance expansion of mutual information for free scalar \CFT{d} is given by
\begin{equation}
I(\xi)\sim\sum_{n=d-2}^{\infty}c_n\,\xi^n,
\end{equation}
with coefficients of the form~\cite{Buchanan:2024tlm}:
\begin{equation}
c_n=
\begin{cases}
q_{n,0},
& d\ \text{even},
\\[2mm]
\displaystyle
\sum_{r=0}^{\lfloor n/(d-2)\rfloor-1}
\frac{q_{n,r}}{\pi^{2r}},
& d\ \text{odd},
\end{cases}
\qquad q_{n,r}\in\mathbb Q.
\end{equation}
Assembling the modular-integral coefficients in~\cref{tab:free-scalar-blocks}, we find
\begin{equation}
\begin{aligned}
d=3:\qquad I(\xi)
&\sim
\frac13\,\xi
+\left(-\frac4{45}+\frac4{3\pi^2}\right)\xi^2
+\left(\frac{13}{21}-\frac8{9\pi^2}+\frac{16}{3\pi^4}\right)\xi^3
+\cdots,
\\[2mm]
d=4:\qquad I(\xi)
&\sim
\frac4{15}\,\xi^2
-\frac{16}{105}\,\xi^3
+\frac87\,\xi^4
-\frac{2672}{1155}\,\xi^5
+\cdots.
\end{aligned}
\end{equation}
These agree with the results in~\cite{Buchanan:2024tlm}, obtained using a method specific
to free fields.

\section{Discussion and outlook}\label{sec:discussion}
\paragraph{Convergence and resummation.}
The good agreement of the truncated expansion in Ising \CFT{2} with lattice data motivates investigating whether the long-distance expansion
converges in this theory. The free compact boson \CFT{2} illustrates a different
possibility: the current contribution has a divergent expansion,
while at the self-dual radius cancellations with vertex operator
contributions produce a convergent result. These examples raise the
question of which physical properties of a CFT determine the
convergence of~\cref{eq:mi-master}, and how cancellations between sectors affect it. A related challenge is to develop
systematic resummation methods that extend the long-distance
expansion to finite separation and ultimately recover the
short-distance behavior of mutual information.

\paragraph{Holographic CFT.} For holographic CFTs in classical gravity approximation, the holographic entanglement entropy formula~\cite{Ryu:2006bv} implies mutual information vanishes to leading order in the long-distance disconnected phase. The long-distance expansion~\cref{eq:mi-master} therefore gives quantum corrections to the classical mutual information, and is expected to be related to bulk entanglement entropy~\cite{Faulkner:2013ana}. The two-point primary contribution~\cref{eq:two-pt-sector-primary} was explained in~\cite{Agon:2015ftl} from bulk perspective; it would be worthwhile to extend their analysis to the all-orders expansion~\cref{eq:mi-master}.

\paragraph{Other information-theoretic quantities.} The radial sewing expansion of density matrix, together with geometric modular flow, also offers a novel approach to other information quantities in \CFT{d}. A proper Rényi generalization of mutual information~\cref{eq:MI-def} is sandwiched Rényi mutual information, where the relative entropy in~\cref{eq:MI-def} is replaced by sandwiched Rényi relative entropy. Reflected entropy is another example, which enjoys geometric representation in holographic \CFT{d}. Application to these quantities will be reported in upcoming works~\cite{CompanionSRMI, CompanionReflected}.

\acknowledgments We acknowledge support from the Tsinghua Shuimu Scholar program, and hospitality from CERN where part of this work was completed. We acknowledge use of ChatGPT and Claude for assistance with derivation, computation and manuscript preparation. We take responsibility for validity of the results.

\appendix
\section{Lattice computation of Ising mutual information}\label{app:ising-lattice}
The critical Ising spin chain has Hamiltonian
\begin{equation}\label{eq:ising-hamiltonian}
H=-\sum_j\left(\sigma_j^x\sigma_{j+1}^x+\sigma_j^z\right),
\end{equation}
where $\sigma_j^\alpha$ are Pauli matrices at site $j$, and the exchange coupling and transverse field are set to unity. We work in the ground state of the infinite chain and consider two disjoint spin intervals $A$ and $B$, containing $\ell_A$ and $\ell_B$ sites and separated by a gap $G$ containing $\ell_G$ sites. Their mutual information and cross ratio are
\begin{equation}
I_{\mathrm{lat}}
=S(\rho_A)+S(\rho_B)-S(\rho_{AB}),
\qquad
\eta=\frac{\ell_A\ell_B}{(\ell_A+\ell_G)(\ell_B+\ell_G)}.
\end{equation}
To compare with Ising \CFT{2}, we increase all three lengths at fixed $\eta$ and extrapolate to the continuum limit.

We evaluate the reduced density matrices using the Jordan–Wigner transformation
\begin{equation}
  a_j=\left(\prod_{k<j}\sigma_k^z\right)\sigma_j^x,
\qquad
b_j=\left(\prod_{k<j}\sigma_k^z\right)\sigma_j^y.
\end{equation}
The Majorana operators bring the Hamiltonian~\cref{eq:ising-hamiltonian} into the quadratic form
\begin{equation}\label{eq:ising-majorana}
H=i\sum_j\left(a_jb_j+b_ja_{j+1}\right).
\end{equation}
The ground state is therefore Gaussian in the Majorana operators. The two-point functions are
\begin{equation}
\langle a_ja_k\rangle=\langle b_jb_k\rangle=\delta_{jk},
\qquad
i\langle a_jb_k\rangle
=\frac{2}{\pi[2(j-k)-1]},
\end{equation}
and higher-point correlators are obtained from Wick’s theorem.

The two-interval reduced density matrix $\rho_{AB}$ is given in terms of Pauli correlators by
\begin{equation}
\rho_{AB}
=
\frac{1}{2^{\ell_A+\ell_B}}
\sum_{\{\mu_j\}}
\left\langle
\prod_{j\in A\cup B}\sigma_j^{\mu_j}
\right\rangle
\prod_{j\in A\cup B}\sigma_j^{\mu_j},
\qquad
\mu_j\in\{0,x,y,z\},
\end{equation}
where $\sigma_j^0=\id$. For each interval $R=A,B$, define the following Majorana monomials
\begin{equation}
O_R(\boldsymbol\mu,\boldsymbol\nu)=\prod_{j\in R}
a_j^{\mu_j}b_j^{\nu_j},
\qquad \mu_j,\nu_j\in\{0,1\},
\end{equation}
where the product is ordered by increasing site index. The reduced density matrix can then be represented in terms of these monomials as~\cite{Fagotti:2010yr}
\begin{equation}
\begin{aligned}
\rho_{AB}
=\frac{1}{2^{\ell_A+\ell_B}}
\biggl[
&\sum_{\mathrm{even}}
\langle O_AO_B\rangle\,
(O_AO_B)^\dagger
+\sum_{\mathrm{odd}}
\langle O_AS_GO_B\rangle\,
(O_AS_GO_B)^\dagger
\biggr],
\end{aligned}
\end{equation}
where $S_G$ is defined by
\begin{equation}
S_G=\prod_{j\in G}\sigma_j^z
=\prod_{j\in G}(-ia_jb_j).
\end{equation}
The sums are over all monomials $O_A$ and $O_B$ with even or odd degree in each interval, respectively.

We construct $\rho_{AB}$ in the product basis of eigenstates of $\rho_A$ and $\rho_B$. For $R=A,B$, denote the correlation matrix as
\begin{equation}
(D_R)_{jk}=i\langle a_jb_k\rangle,
\qquad j,k\in R.
\end{equation}
The matrix has singular value decomposition
\begin{equation}
D_R=U_R\,\operatorname{diag}(s_{R,k})\,V_R^{\mathsf T}.
\end{equation}
Let $r$ be the leftmost site of $R$, and we define
\begin{equation}
a_j^{(R)}
=
\left(\prod_{r\le k<j}\sigma_k^z\right)\sigma_j^x,
\qquad
b_j^{(R)}
=
\left(\prod_{r\le k<j}\sigma_k^z\right)\sigma_j^y.
\end{equation}
Their two-point matrix is given by the same $D_R$. We define the following fermionic modes
\begin{equation}\label{eq:f-mode}
f_{R,k}
=
\frac12\sum_{j\in R}
\left[
(U_R)_{jk}a_j^{(R)}
-i(V_R)_{jk}b_j^{(R)}
\right].
\end{equation}
They obey the canonical anti-commutation relations due to orthogonality of $U_R,V_R$. Their two-point functions are
\begin{equation}
\langle f_{R,k}^{\dagger}f_{R,l}\rangle
=\delta_{kl}\frac{1-s_{R,k}}2,
\qquad
\langle f_{R,k}f_{R,l}\rangle=0.
\end{equation}
Now define the normalized mode vacuum and occupation states by
\begin{equation}
f_{R,k}|0;R\rangle=0,
\qquad
|\boldsymbol n;R\rangle
=
(f_{R,1}^{\dagger})^{n_1}\cdots
(f_{R,\ell_R}^{\dagger})^{n_{\ell_R}}
|0;R\rangle, \qquad n_k\in\{0,1\}.
\end{equation}
Since $\rho_R$ is Gaussian, these occupation states diagonalize it
\begin{equation}
\rho_R|\boldsymbol n;R\rangle
=p_R(\boldsymbol n)|\boldsymbol n;R\rangle,
\end{equation}
with eigenvalues
\begin{equation}
p_R(\boldsymbol n)
=
\prod_{k=1}^{\ell_R}
\left(\frac{1+s_{R,k}}2\right)^{1-n_k}
\left(\frac{1-s_{R,k}}2\right)^{n_k}.
\end{equation}

For each interval $R$, we select a set $\mathcal K_R$ of occupation states with the largest probabilities $p_R(\boldsymbol n)$. We denote the matrix element of $\rho_{AB}$ in the restricted product basis as
\begin{equation}
\bigl(\rho_{AB}^{\mathrm{ret}}\bigr)_{\boldsymbol n_A\boldsymbol n_B;\boldsymbol m_A\boldsymbol m_B}
    =
\langle\boldsymbol n_A;A|
\langle\boldsymbol n_B;B|
\rho_{AB}
|\boldsymbol m_A;A\rangle
|\boldsymbol m_B;B\rangle,
\end{equation}
where states associated with each interval belong to $\mathcal K_R$. The matrix element can be computed as follows. We define the following operator
\begin{equation}\label{eq:E-operator}
\begin{aligned}
E_R(\boldsymbol m,\boldsymbol n)
&=|\boldsymbol m;R\rangle\langle\boldsymbol n;R| =
\left[\prod_{k=1}^{\ell_R}(f_{R,k}^{\dagger})^{m_k}\right]
\left[\prod_{k=1}^{\ell_R}f_{R,k}f_{R,k}^{\dagger}\right]
\left[\prod_{k=\ell_R}^{1}f_{R,k}^{n_k}\right],
\end{aligned}
\end{equation}
where the middle factor is the projector $|0;R\rangle\langle0;R|$. The matrix element is then given by
\begin{equation}\label{eq:R-E-2pt}
\bigl(\rho_{AB}^{\mathrm{ret}}\bigr)_{\boldsymbol n_A\boldsymbol n_B;\boldsymbol m_A\boldsymbol m_B}
=
\left\langle
E_A(\boldsymbol m_A,\boldsymbol n_A)
E_B(\boldsymbol m_B,\boldsymbol n_B)
\right\rangle.
\end{equation}
We now define a global counterpart of the modes in~\cref{eq:f-mode} by
\begin{equation}
\widetilde f_{R,k}
=
\frac12\sum_{j\in R}
\left[(U_R)_{jk}a_j-i(V_R)_{jk}b_j\right],
\end{equation}
and similarly $\widetilde E_R$ by replacing every $f_{R,k}$ in~\cref{eq:E-operator} by $\widetilde f_{R,k}$. For the matrix element under consideration, define
\begin{equation}
\epsilon_R=
\left[\sum_k(m_{R,k}+n_{R,k})\right]\bmod2.
\end{equation}
The two-point function in~\cref{eq:R-E-2pt} can be computed from these global operators as
\begin{equation}
\left\langle E_AE_B\right\rangle
=
\begin{cases}
\left\langle\widetilde E_A\widetilde E_B\right\rangle,
&\epsilon_A=\epsilon_B=0,\\[1mm]
\left\langle\widetilde E_A S_AS_G\widetilde E_B\right\rangle,
&\epsilon_A=\epsilon_B=1,\\[1mm]
0,&\epsilon_A\ne\epsilon_B.
\end{cases}
\end{equation}
Now everything is explicitly built from global Majorana operators. In particular,
\begin{equation}
S_AS_G
=
(-i)^{\ell_A+\ell_G}
\prod_{j\in A\cup G} a_jb_j.
\end{equation}

Together with the single interval entropies fixed by $p_R(\boldsymbol n)$, we then compute the mutual information from the restricted density matrix $\rho_{AB}^{\mathrm{ret}}$ as
\begin{equation}\label{eq:ising-compressed-mi}
\begin{aligned}
I_{\mathrm{comp}}
={}&
\sum_{R=A,B}S(\rho_R^{\mathrm{ret}}\oplus\delta_R)
-S\Bigl[
\rho_{AB}^{\mathrm{ret}}
\oplus
\bigl(\rho_A^{\mathrm{ret}}
-\operatorname{Tr}_B\rho_{AB}^{\mathrm{ret}}\bigr)
\\
&\qquad
\oplus
\bigl(\rho_B^{\mathrm{ret}}
-\operatorname{Tr}_A\rho_{AB}^{\mathrm{ret}}\bigr)
\oplus
\bigl(\delta_A+\delta_B-1
+\operatorname{Tr}\rho_{AB}^{\mathrm{ret}}\bigr)
\Bigr].
\end{aligned}
\end{equation}
Here $\rho_R^{\mathrm{ret}}$ is the diagonal matrix of probabilities $p_R(\boldsymbol n)$ with $\boldsymbol n\in\mathcal K_R$, and $\delta_R=1-\operatorname{Tr}\rho_R^{\mathrm{ret}}$ is the discarded probability. Its difference with the full mutual information $I_{\mathrm{lat}}$ is bounded by
\begin{equation}
0\le I_{\mathrm{lat}}-I_{\mathrm{comp}}\le \mathcal{E}_A+\mathcal{E}_B,
\end{equation}
where
\begin{equation}
\mathcal{E}_R=
h_2(\delta_R)
+
2\left[
S(\rho_R)
+\sum_{\boldsymbol n\in\mathcal K_R}
p_R(\boldsymbol n)\log p_R(\boldsymbol n)
+\delta_R\log\delta_R
\right],
\end{equation}
$h_2(x)=-x\log x-(1-x)\log(1-x)$ is the binary entropy function.
At each fixed lattice geometry, we enlarge $\mathcal K_A$ and $\mathcal K_B$ until the basis-truncation error in mutual information is below $5\times10^{-6}$. We increase the interval lengths and gap at fixed $\eta$ and extrapolate $I_{\mathrm{lat}}$ to the continuum. We assess the extrapolation uncertainty by varying the fitting formula and the sizes included.

\FloatBarrier
\printbibheading
\printbibliography[heading=none]

\end{document}

%% file: macros.tex
\newcommand{\prn}[1]{\left( #1 \right)}

\newcommand{\abs}[1]{\left| #1 \right|}

\newcommand{\CFT}[1]{\ensuremath{\mathrm{CFT}_{#1}}}

\newcommand{\id}{\ensuremath{\mathds{1}}}

%% file: figures/radial-expansion.tex
\begingroup
\definecolor{radialpink}{HTML}{CC5687}
\definecolor{radialblue}{HTML}{427EAE}
\definecolor{radialgold}{HTML}{D5B65A}
\begin{tikzpicture}[
  x=1cm,y=1cm,font=\small,
  frame/.style={draw=black,line width=0.55pt},
  sphere/.style={draw=black,line width=0.65pt},
  cut/.style={line width=1pt,line cap=round},
  every node/.style={inner sep=2pt}
]
\node[anchor=east,font=\normalsize] at (0.78,2.55) {$\rho_{AB}$};
\node[font=\normalsize] at (1.15,2.55) {$=$};
\node[font=\normalsize] at (2.15,2.55) {$\displaystyle\sum_{a,b}$};
\fill[radialblue!12] (3.10,0.55) rectangle (9.70,4.55);
\fill[radialgold!18] (6.40,2.55) circle[radius=1.70];
\fill[radialpink!12] (6.40,2.55) circle[radius=0.82];
\draw[frame] (3.10,0.55) rectangle (9.70,4.55);
\draw[sphere] (6.40,2.55) circle[radius=1.70];
\draw[sphere] (6.40,2.55) circle[radius=0.82];
\node[anchor=west,font=\footnotesize] at (7.90,1.10) {$|x|=1$};
\node[font=\footnotesize] at (6.40,3.65) {$|x|=\xi$};
\node[anchor=east] at (5.01,3.62) {$|b\rangle$};
\node[anchor=east] at (5.56,2.95) {$\langle a|$};
\draw[cut,radialpink] plot[domain=5.58:7.22,samples=81]
  (\x,{2.55+0.035*sin(5*360*(\x-5.58)/1.64)});
\draw[cut,radialblue] plot[domain=3.10:4.70,samples=81]
  (\x,{2.55+0.035*sin(5*360*(\x-3.10)/1.60)});
\draw[cut,radialblue] plot[domain=8.10:9.70,samples=81]
  (\x,{2.55+0.035*sin(5*360*(\x-8.10)/1.60)});
\node[text=radialpink] at (6.40,2.88) {$A$};
\node[text=radialblue] at (8.90,2.88) {$B$};
\node[font=\footnotesize] at (6.40,2.13) {$\mathscr R_A$};
\node[font=\footnotesize] at (8.90,1.85) {$\mathscr R_B$};
\node[font=\footnotesize] at (6.40,1.30) {$\mathscr R_{\mathrm{ann}}$};
\draw[->,>=stealth,line width=0.5pt] (8.65,3.65) -- (8.65,4.19)
  node[above,font=\footnotesize] {$\tau$};
\draw[->,>=stealth,line width=0.5pt] (8.65,3.65) -- (9.20,3.65)
  node[right,font=\footnotesize] {$x^1$};

\node[font=\normalsize] at (1.15,-1.27) {$=$};
\node[anchor=west,font=\normalsize] at (1.67,-1.27)
  {$\displaystyle\sum_{a,b}\xi^{\Delta_a}(G^{-1})_{ba}$};
\draw[frame,fill=radialpink!12] (5.05,-2.35) rectangle (8.82,-0.19);
\draw[frame,fill=radialblue!12] (9.69,-2.35) rectangle (13.46,-0.19);
\node[font=\normalsize] at (9.25,-1.27) {$\otimes$};

\draw[cut,radialpink] plot[domain=5.70:6.93,samples=81]
  (\x,{-1.27+0.035*sin(4*360*(\x-5.70)/1.23)});
\node[text=radialpink] at (6.31,-0.94) {$A$};
\fill[radialpink] (7.84,-1.26) circle[radius=2pt];
\node[anchor=south] at (7.84,-1.03) {$\mathcal O_{\bar a}(\infty)$};
\node[anchor=north] at (6.94,-2.45) {$h_{A,a}$};

\draw[cut,radialblue] plot[domain=9.69:10.70,samples=61]
  (\x,{-1.27+0.035*sin(3*360*(\x-9.69)/1.01)});
\draw[cut,radialblue] plot[domain=12.45:13.46,samples=61]
  (\x,{-1.27+0.035*sin(3*360*(\x-12.45)/1.01)});
\node[text=radialblue] at (12.96,-0.94) {$B$};
\fill[radialblue] (11.58,-1.27) circle[radius=2pt];
\node[anchor=south] at (11.58,-1.03) {$\mathcal O_b(0)$};
\node[anchor=north] at (11.58,-2.45) {$h_{B,b}$};
\end{tikzpicture}
\endgroup

%% file: figures/modular-cap-preparation.tex
\begingroup
\usetikzlibrary{decorations.markings,arrows.meta}
\definecolor{preppink}{HTML}{CC5687}
\definecolor{prepblue}{HTML}{427EAE}
\begin{tikzpicture}[
  x=1cm,y=1cm,font=\small,
  frame/.style={draw=black,line width=0.55pt},
  cut/.style={line width=1pt,line cap=round},
  evolution/.style={
    draw=black,line width=0.65pt,
    postaction={decorate},
    decoration={markings,mark=at position 0.5 with
      {\arrow{Stealth[length=4pt,width=3.5pt]}}}
  },
  every node/.style={inner sep=2pt}
]
\begin{scope}[yshift=3.4cm]
  \draw[frame,fill=preppink!12] (0,-1.05) rectangle (3.45,1.05);
  \draw[cut,preppink] plot[domain=0.48:1.72,samples=65]
    (\x,{0.035*sin(4*360*(\x-0.48)/1.24)});
  \node[text=preppink] at (1.10,0.34) {$A$};
  \fill[preppink] (2.65,0) circle[radius=2pt];
  \node[anchor=south] at (2.65,0.20) {$\mathcal O_{\bar a}(\infty)$};
  \node[anchor=north] at (1.725,-1.13) {$h_{A,a}$};
  \node[font=\normalsize] at (3.87,0) {$=$};

  \begin{scope}[xshift=6.75cm]
    \fill[preppink!12] (-0.92,-1.20) rectangle (0.92,1.20);
    \fill[preppink!12] (0,1.20) ellipse[x radius=0.92,y radius=0.26];
    \fill[preppink!12] (0,-1.20) ellipse[x radius=0.92,y radius=0.26];
    \draw[frame] (0.92,-1.20) -- (0.92,1.20);
    \draw[frame] (-0.92,1.20)
      arc[start angle=180,end angle=-180,x radius=0.92,y radius=0.26];
    \draw[frame] (-0.92,-1.20)
      arc[start angle=180,end angle=-180,x radius=0.92,y radius=0.26];
    \draw[cut,preppink] plot[domain=-1.2:1.2,samples=71]
      ({-0.92+0.016*sin(4*360*(\x+1.2)/2.4)},\x);
    \node[anchor=east,font=\footnotesize] at (-1.04,1.35)
      {$\varphi=0,2\pi$};

    \draw[evolution] (-0.92,0)
      .. controls (-0.60,-0.43) and (0.60,-0.43) .. (0.92,0);
    \draw[evolution,densely dashed] (0.92,0)
      .. controls (0.60,0.43) and (-0.60,0.43) .. (-0.92,0);
    \node at (0,-0.64) {$\rho_A^{1/2}$};
    \node at (0,0.64) {$\rho_A^{1/2}$};
    \fill[preppink] (0.92,0) circle[radius=2pt];
    \node[anchor=west] at (1.07,0.14) {$\mathcal O_{\bar a}(\infty)$};
    \node[anchor=west,font=\footnotesize] at (1.07,-0.29)
      {$\varphi=\pi$};
  \end{scope}
  \node[font=\normalsize] at (9.85,0) {$\propto$};
  \node[anchor=west,font=\normalsize] at (10.30,0)
    {$\rho_A^{1/2}\mathcal O_a^\dagger(0)\rho_A^{1/2}$};
\end{scope}

\begin{scope}
  \draw[frame,fill=prepblue!12] (0,-1.05) rectangle (3.45,1.05);
  \draw[cut,prepblue] plot[domain=0:0.92,samples=49]
    (\x,{0.035*sin(3*360*\x/0.92)});
  \draw[cut,prepblue] plot[domain=2.53:3.45,samples=49]
    (\x,{0.035*sin(3*360*(\x-2.53)/0.92)});
  \node[text=prepblue] at (2.99,0.34) {$B$};
  \fill[prepblue] (1.725,0) circle[radius=2pt];
  \node[anchor=south] at (1.725,0.20) {$\mathcal O_b(0)$};
  \node[anchor=north] at (1.725,-1.13) {$h_{B,b}$};
  \node[font=\normalsize] at (3.87,0) {$=$};

  \begin{scope}[xshift=6.75cm]
    \fill[prepblue!12] (-0.92,-1.20) rectangle (0.92,1.20);
    \fill[prepblue!12] (0,1.20) ellipse[x radius=0.92,y radius=0.26];
    \fill[prepblue!12] (0,-1.20) ellipse[x radius=0.92,y radius=0.26];
    \draw[frame] (-0.92,-1.20) -- (-0.92,1.20);
    \draw[frame] (0.92,1.20)
      arc[start angle=0,end angle=360,x radius=0.92,y radius=0.26];
    \draw[frame] (0.92,-1.20)
      arc[start angle=0,end angle=360,x radius=0.92,y radius=0.26];
    \draw[cut,prepblue] plot[domain=-1.2:1.2,samples=71]
      ({0.92+0.016*sin(4*360*(\x+1.2)/2.4)},\x);
    \node[anchor=west,font=\footnotesize] at (1.04,1.35)
      {$\varphi=\pi,-\pi$};

    \draw[evolution] (0.92,0)
      .. controls (0.60,-0.43) and (-0.60,-0.43) .. (-0.92,0);
    \draw[evolution,densely dashed] (-0.92,0)
      .. controls (-0.60,0.43) and (0.60,0.43) .. (0.92,0);
    \node at (0,-0.64) {$\rho_B^{1/2}$};
    \node at (0,0.64) {$\rho_B^{1/2}$};
    \fill[prepblue] (-0.92,0) circle[radius=2pt];
    \node[anchor=east] at (-1.07,0.14) {$\mathcal O_b(0)$};
    \node[anchor=east,font=\footnotesize] at (-1.07,-0.29)
      {$\varphi=0$};
  \end{scope}
  \node[font=\normalsize] at (9.85,0) {$\propto$};
  \node[anchor=west,font=\normalsize] at (10.30,0)
    {$\rho_B^{1/2}\mathcal O_b^{(-\pi)}(0)\rho_B^{1/2}$};
\end{scope}
\end{tikzpicture}
\endgroup

%% file: figures/lorentzian-modular-flow.tex
\begingroup
\definecolor{cappink}{HTML}{CC5687}
\definecolor{capblue}{HTML}{427EAE}
\begin{tikzpicture}[
  x=1.8cm,y=1.8cm,
  font=\small,
  boundary/.style={draw=black!60,line width=0.65pt},
  orbit/.style={draw=black!65,line width=0.8pt,->,>=stealth},
  field/.style={inner sep=2.5pt},
  guide/.style={draw=black!35,densely dashed,line width=0.4pt}
]
\pgfmathsetmacro{\halfcircle}{pi/2}
\pgfmathsetmacro{\fullcircle}{pi}
\pgfmathsetmacro{\rightedge}{3*pi/2}
\pgfmathsetmacro{\capheight}{2*atan(1/4)*pi/180}

\path[boundary,fill=cappink!12] (-\halfcircle,0) -- (0,\halfcircle)
  -- (\halfcircle,0) -- (0,-\halfcircle) -- cycle;
\path[boundary,fill=capblue!12] (\halfcircle,0) -- (\fullcircle,\halfcircle)
  -- (\rightedge,0) -- (\fullcircle,-\halfcircle) -- cycle;

\draw[guide] (-\halfcircle,-1.65) -- (-\halfcircle,1.65);
\draw[guide] (\rightedge,-1.65) -- (\rightedge,1.65);
\node[text=black!55] at (-\halfcircle,1.80) {$\sim$};
\node[text=black!55] at (\rightedge,1.80) {$\sim$};

\draw[orbit] (0,-1.40) -- (0,1.40);
\draw[orbit] (\fullcircle,-1.40) -- (\fullcircle,1.40);
\node[anchor=east,inner sep=3pt] at (0,1.03) {$U_{is}$};
\node[anchor=west,inner sep=3pt] at (\fullcircle,1.03) {$U_{-is}$};
\draw[orbit] (\halfcircle,1.18) -- (\halfcircle,1.76)
  node[above] {$T$};
\draw[orbit] (\halfcircle,1.18) -- ({\halfcircle+0.58},1.18)
  node[right] {$\chi$};

\fill[cappink] (0,-\capheight) circle[radius=2pt];
\fill[cappink] (0,\capheight) circle[radius=2pt];
\fill[cappink] (\fullcircle,0) circle[radius=2pt];
\node[field,anchor=west] at (0,-\capheight)
  {$(\mathcal O_{a_1}^{\dagger})^{(is_1)}$};
\node[field,anchor=west] at (0,\capheight)
  {$(\mathcal O_{a_2}^{\dagger})^{(is_2)}$};
\node[field,anchor=west] at (\fullcircle,0)
  {$(\mathcal O_{a_0}^{\dagger})^{(\pi)}$};

\fill[capblue] (0,0) circle[radius=2pt];
\fill[capblue] (\fullcircle,-\capheight) circle[radius=2pt];
\fill[capblue] (\fullcircle,\capheight) circle[radius=2pt];
\node[field,anchor=east] at (0,0) {$\mathcal O_{b_0}$};
\node[field,anchor=east] at (\fullcircle,-\capheight)
  {$\mathcal O_{b_1}^{(\pi-is_1)}$};
\node[field,anchor=east] at (\fullcircle,\capheight)
  {$\mathcal O_{b_2}^{(\pi-is_2)}$};

\node[align=center] at (0,-1.89)
  {$D(\mathcal B_1)$};
\node[align=center] at (\fullcircle,-1.89)
  {$D(\mathcal B_1^c)$};

\fill[cappink] (-0.72,-2.35) circle[radius=2pt];
\node[anchor=west,inner sep=4pt] at (-0.72,-2.35)
  {inner ball factor};

\fill[capblue] ({\fullcircle-0.72},-2.35) circle[radius=2pt];
\node[anchor=west,inner sep=4pt] at ({\fullcircle-0.72},-2.35)
  {outer ball factor};
\end{tikzpicture}
\endgroup